\pdfoutput=1
\documentclass{article}
\usepackage[utf8]{inputenc}
\usepackage{url, hyperref}
\usepackage{amssymb, amsfonts, amsmath}
\usepackage{graphicx, natbib}
\usepackage{algorithm2e}
\usepackage{authblk}
\newcommand{\BIT}{\begin{itemize}}
\newcommand{\EIT}{\end{itemize}}
\newcommand{\BNUM}{\begin{enumerate}}
\newcommand{\ENUM}{\end{enumerate}}

\def\reals{\mathbb{R}} 

\def\Gsn{\mathcal{N}}

\def\Mult{\textnormal{Mult}}

\def\*#1{\mathbf{#1}}

\title{Interactive Visualization of Spatial Omics Neighborhoods}
\author{Tinghui Xu and Kris Sankaran}
\affil{Department of Statistics \\ 
University of Wisconsin - Madison
}

\begin{document}

\maketitle

\begin{abstract}
Dimensionality reduction of spatial omic data can reveal shared, spatially structured patterns of expression across a collection of genomic features. We study strategies for discovering and interactively visualizing low-dimensional structure in spatial omic data based on the construction of neighborhood features. We design quantile and network-based spatial features that result in spatially consistent embeddings. A simulation compares embeddings made with and without neighborhood-based featurization, and a re-analysis of \citep{keren2019mibi} illustrates the overall workflow. We provide an R package, NBFvis, to support computation and interactive visualization for the proposed dimensionality reduction approach. Code and data for reproducing experiments and analysis is available at \url{https://github.com/XTH1114/NBFvis}.
\end{abstract}

Spatially resolved omic technologies provide a view into the landscape of complex biological processes \citep{ burgess2019spatial,nawy2018spatial}. For example, they have revealed novel aspects of tissue differentiation and the structure of certain cancers \citep{rao2021exploring, yoosuf2020identification}. A spatial transcriptomic or proteomic dataset can be viewed as a spatially indexed collection of high-dimensional vectors \citep{dries2021advances}. The coordinates of each vector correspond to different genomic features (genes expression and protein measurements for spatial transcriptomics and proteomics, respectively) while the spatial index locates each measurement at some location in a tissue or cell.

Two challenges in the analysis of spatial omic data are,
\begin{itemize}
\item Microenvironment dimensionality reduction: Considering the large number of simultaneously measured genomic features, some form of dimensionality reduction is essential for effective exploratory analysis. However, for spatially resolved data, a dimensionality reduction should describe microenvironments and their relationships with one another. It is more useful to embed the genomic signature of a cell’s local neighborhood than simply the cell in isolation.
\item Streamlined navigation: Low-dimensional representations of microenvironments may not be interpretable on their own. To this end, it is helpful to relate the representations the  to their original spatial and genomic contexts. Ensuring that these correspondences can be explored efficiently is a challenge in itself.
\end{itemize}

This paper discusses methods to address these challenges and releases a new R package that implements them. For first challenge, our approach is to featurize spatial neighborhoods and pass this representation to downstream dimensionality reduction techniques. We explore in depth features based on (1) histograms of expression levels and (2) local cell network properties. For the second challenge, we design an interactive visualization that links learned representations with contextual descriptors.

We evaluate these methods using simulation and a qualitative data analysis. The simulation clarifies the difference between learning representations on individual cells and local cellular neighborhoods. The data analysis recapitulates the findings of \citep{chen2020modeling, keren2019mibi}. We believe that the main advantages of the proposed approach are,
\begin{itemize}
\item Modularity: The approach can be make use of existing dimensionality-reduction methods while ensuring that results reflect meaningful spatial structure. 
\item Flexibility: Spatial featurizations can be tailored to specific problem contexts with little changes to the overall workflow.
\end{itemize}
Our methods are implemented in the R package NBFvis, available at \url{https://github.com/XTH1114/NBFvis}.

The remainder of the paper is organized as follows. Section \ref{sec:background} reviews relevant literature on analysis of spatial omic data. Section \ref{sec:methods} describes the proposed method. Section \ref{sec:visualization} introduces what kinds of visualization and interactivity are provided in our package. Sections \ref{sec:simulation} and \ref{sec:data} illustrate the method in simulation and real data analysis. Section \ref{sec:package} gives an overview of NBFVis’s functionality. We conclude with a summary and directions for future work in Section \ref{sec:discussion}.

\section{Background}
\label{sec:background}

The proliferation of spatial omic data has attracted attention from the modeling and visualization communities. Important themes that have emerged include the selection of spatially varying genes, derivation of spatial summary measures, and discovery of spatially consistent microenvironments. The resulting software packages allow analysts to generate overviews of spatial variation as well as focus on specific genomic features of interest.

Several studies propose feature-level models of spatial variation to select those with notable spatial expression patterns. SPARK fits a collection of random effects models with diverse set of kernels to capture variation at several spatial scales \cite{sun2020statistical}. Alternatively, \citep{zhu2020integrative} computes a measure of spatial variation based on a spatially induced graph laplacian; genes exhibiting similar patterns of spatial expression are then clustered. Alternatively, \citep{hsubirs} proposes an adaptation of Moran’s $I$-statistic to measure the extent of spatial clustering across cell types, highlighting the potential for the classical spatial statistics methods to support modern spatial omics analysis. Like NBFVis, these methods compute spatial statistics to summarize spatial omic datasets. However, they tend not to provide localized measures of spatial structure, focusing instead on tissue-level properties.

The Giotto package includes approaches to dimensionality reduction and interactive visualization of spatial omics data \cite{dries2021giotto}. Of particular interest, the package supports interactive visualization that dynamically links embeddings of expression measurements with corresponding cell locations. Note however that these embeddings are derived without reference to spatial features.

Similar to our approach, Spatial-LDA proposes a variation of the topic models that learns spatially consistent patterns of cell type mixing \citep{chen2020modeling}. This is achieved by tying together mixed memberships of neighboring cells in a structured prior, and the model is fit using a custom optimization scheme. Regions with similar topic memberships can be interpreted as microenvironments. Our proposal has a similar data analytic goal; however, we aim to support more generic spatial features while preserving simplicity in implementation.

\section{Simulation}
\label{sec:simulation}

We provide a toy simulation to clarify the differences in embeddings when neighborhood information is and is not used. We find that if only cell-level information is considered, the embeddings will be dominated by cell types and fail to reflect microenvironment structure.

\subsection{Dataset construction}

Assume that there is one tissue section with three cell types. For each cell, five proteins are measured. Different cell types have different protein profiles, which means the average measurements of proteins differs according to cell type. We assume that cell types are clustered spatially, but that these clusters are close enough so that some areas overlap. These overlapping areas can be considered different microenvironments, since the local mixture of protein profiles is different from regions of pure cell types.

\begin{figure}
    \centering
    \includegraphics[width=1\textwidth]{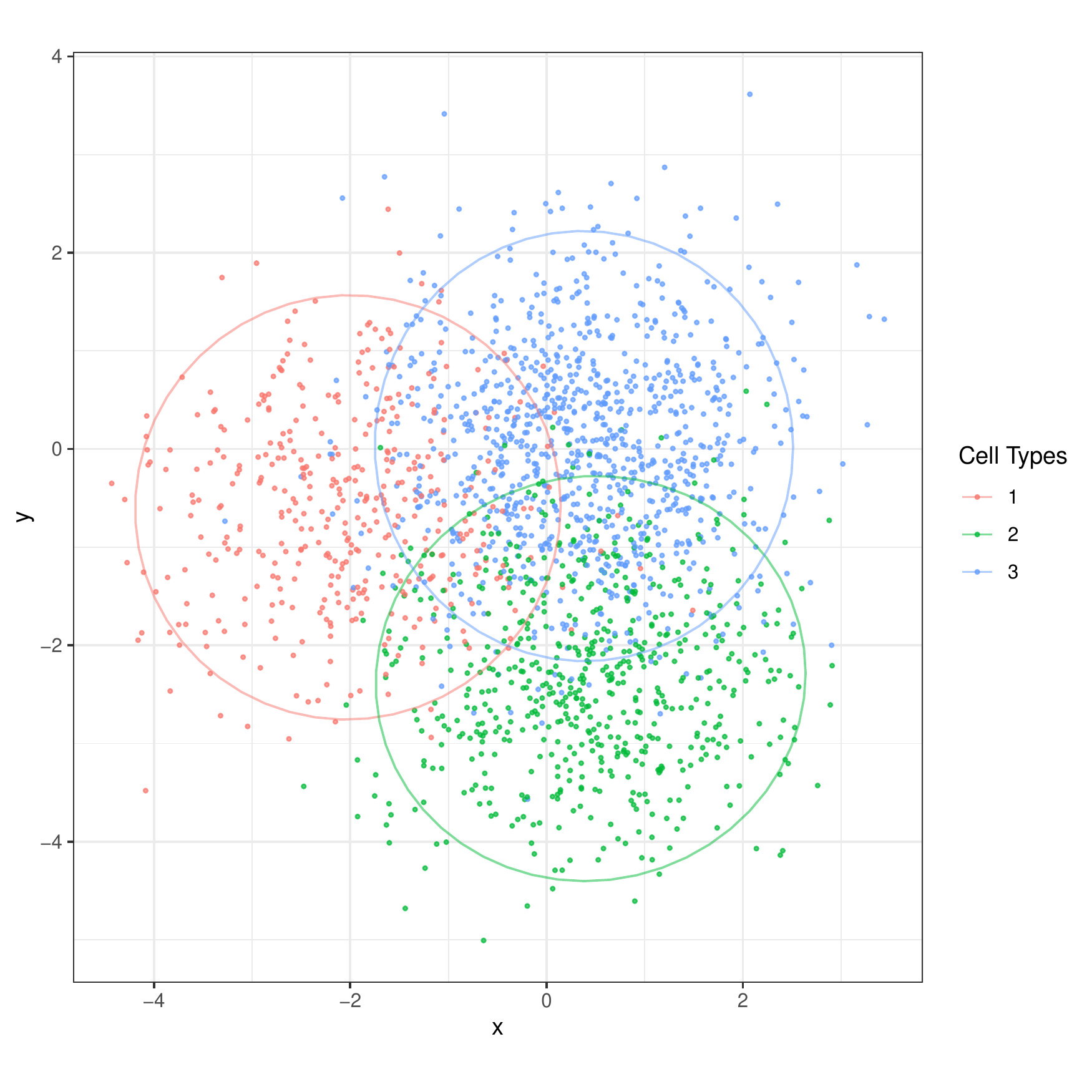}
    \caption{Simulation of a tissue section with three different, partially overlapping cell types. Overlapping regions can be thought of as their distinct microenvironments. The goal is construct embeddings that reflect different mixing patterns.}
    \label{fig:simulation data}
\end{figure}

Figure \ref{fig:simulation data} is the spatial plot for the simulated dataset. Two thousand ``cells'' are generated and divided into three different cell types according to a multinomial distribution with the probability $(0.2, 0.3, 0.5)$ for Cell Type 1, 2, and 3,
\begin{align*}
c_{i}\sim \Mult\left(2000,\left(0.2,0.3,0.5\right)\right), i = 1, \dots, 2000,
\end{align*}
where $c_i$ is the cell type of the $i^{th}$ cell.

For this demonstration, we imagine that protein abundances are drawn from a mixture of multivariate normals. The average of each mixture component represents the typical cell profile for each cell type. That is, for each cell, the measurements for each of the five proteins has the form,
\begin{align*}
p_{i} \vert \mu_{c_{i}} &\sim \Gsn(\mu_{c_i},5 I_5), i=1,\dots,2000 \\
\mu_j &\sim \Gsn\left(0,8 I_5\right), j = 1, 2, 3,
\end{align*}
where $\mu_j$ is the average protein profile for the $j^{th}$ cell type and
$p_{i}$ is 5-dimensional measurement for the $i^{th}$ cell.

Next, we simulate cell locations to get mixed spatial patterns. We use a different mixture of (now two-dimensional) multivariate normals. As before, component means $\text{center}_1, \text{center}_2,$ and $\text{center}_3$ are drawn from a multivariate normal. Denoting the coordinates of cell $i$ by $\left(x_i, y_i\right)$ and the spatial mean of cell type $j$ by $\text{center}_{j}$, we draw,
\begin{align*}
(x_{i}, y_{i}) \vert \text{center}_{c_i} &\sim \Gsn\left(\text{center}_{c_i}, 2I_{2}\right)\\
\text{center}_j&\sim \Gsn\left(0, 10 I_{2}\right).
\end{align*}

After simulation, we obtain a $2000\times5$ expression matrix, each row of which corresponds the simulated observation of one cell. We call this matrix the ``single cell matrix.''

To extract neighborhood information for each cell, we first find neighborhoods with a given radius (here we use 0.2 units in length). We then calculate statistics within each neighborhood. We choose quantiles of protein content as neighborhood-based statistics, which are simple but effective. For every cell and protein, we derive 21 quantiles $q_0, q_{0.05}, ..., q_1$ in the neighborhood. After calculation, an extended $2000\times105$ matrix is obtained. We call this the ``neighborhood matrix.''

\subsection{Comparison}

We next apply dimensionality reduction methods to both the single cell and the neighborhood matrices, in order to clarify the difference in the resulting embeddings.

\begin{figure}
    \centering
    \includegraphics[width=1\textwidth]{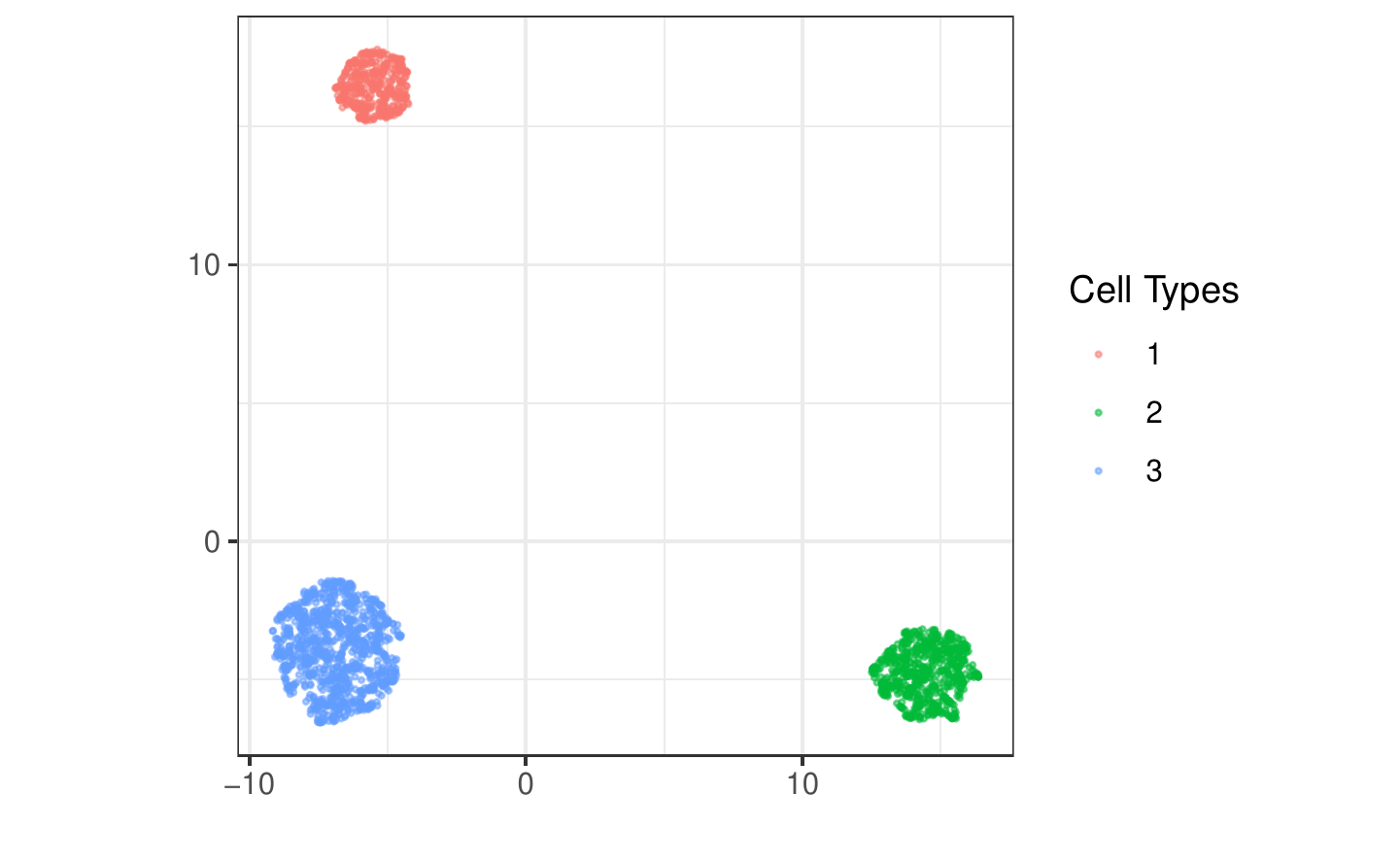}
    \caption{A UMAP embedding plot on the single cell matrix. Cell types cluster with one another, but different mixture patterns are not observed. The embeddings are dominated by the cell types, obscuring the presence of microenvironments.}
    \label{fig:simulation single cell}
\end{figure}

First, we apply Uniform Manifold Approximation and Projection (UMAP)  \cite{mcinnes2018umap} to the single cell matrix, whose low-dimensional embeddings are shown in Figure \ref{fig:simulation single cell}. Only three separate clusters are visible in the embedding plot, each corresponding to a cell type. These UMAP embeddings ignore the microenvironments of mixed cell types along cluster borders in the spatial plot. This result indicates that, when spatial information is not directly incorporated, the low-dimensional embeddings are dominated by cell types and fail to distinguish microenvironments.

\begin{figure}
    \centering
    \includegraphics[width=1\textwidth]{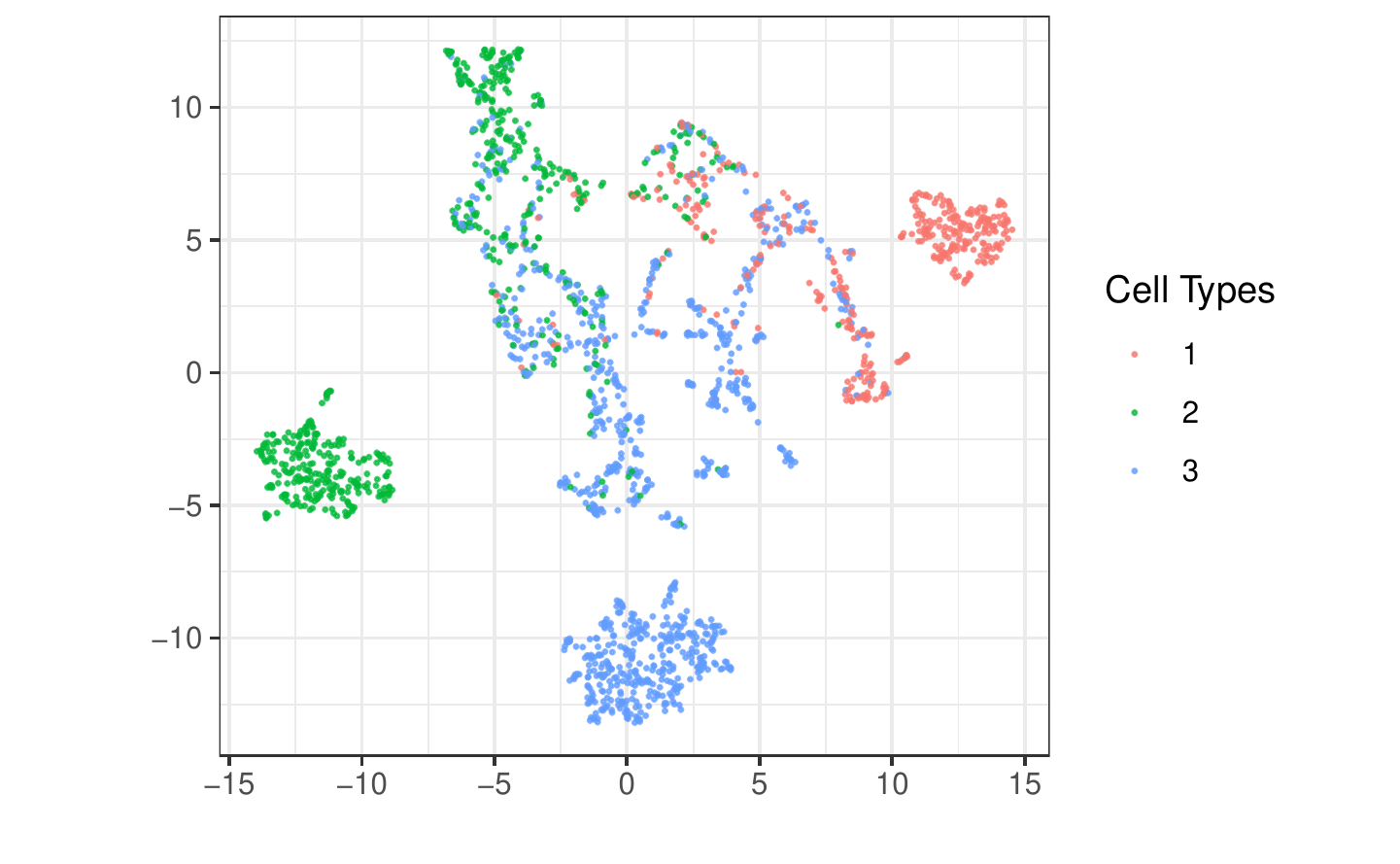}
    \caption{A UMAP embedding plot on the neighborhood matrix. Though simple, the neighborhood quantile statistics make it possible to detect mixture microenvironments. We could further find subclusters like the red-green mixture in the central cluster.}
    \label{fig:simulation neighborhood}
\end{figure}

In contrast, the embedding plot of the neighborhood matrix detects microenvironment structure; see Figure \ref{fig:simulation neighborhood}. We notice that there are still three clusters consisting of pure cell types. However, there are additional clusters of mixed cell types. Between the three pure clusters is a region corresponding to microenvironments with mixed cell types in the spatial plot. Furthermore, we notice that this region can be further divided into spatially consistent ``subcluters.'' For instance, one region with only blue and green cells is related to the blue-green spatial boundary. This can be treated as a unique microenvironment. Similar red-blue and red-green regions are also visible.

In summary, UMAP embeddings using the single cell matrix are dominated by cell types and fail to detect microenvironments with mixed cell types. However, by simply applying UMAP to the neighborhood matrix, we are able to detect these spatially meaningful microenvironments.

\section{Methods}
\label{sec:methods}

First we establish notation and overview the general approach. Let $X \in \reals^{N \times D}$ contain expression measurements for $D$ gene or protein expression features across $N$ cells. We call $X$ the expression matrix.
Let $s \in \reals^{N \times 2}$ contain the spatial locations of the $N$ cells.
We first apply a preliminary dimensionality reduction, like Principal Component Analysis (PCA), to the expression matrix $X$ before the following neighborhood-based featurization. Call the reduced matrix $\hat{X} \in \reals^{N \times P}$, where $P$ is the number of dimensions after dimensionality reduction.

Before we can embed properties of cell neighborhoods, we need to define and derive features for each neighborhood. For each cell $x_{i}$, we define its neighborhood using distances induced by $s$, either containing all cells within a certain radius or simply the $K$-nearest neighbors. Denote the neighborhood for cell $i$ by $m\left(i\right) = \left(x_1, \dots, x_{n_i}\right)$, where $x_1,...x_{n_i}$ where $n_i$ is the number of neighbors surrounding $x_i$.
We featurize the neighborhoods $m\left(i\right)$ using neighborhood-based featurizaton functions $T_j$,$j = 1,...,J$. We then rescale all the derived featurization matrices $T_1(\hat{X}), T_2(\hat{X}),...,T_J(\hat{X})$ and concatenate them to obtain an extended neighborhood-based featurization $T(X)$. The neighborhood matrix from Section \ref{sec:simulation} is a special case of $T\left(X\right)$ using quantile features.

In more detail, let $T_j : \reals^{P} \to \reals^{p_j}, j = 1,2,...,J$ be a set of featurization functions. By applying every $T_j\left(m\left(i\right)\right)$ to each neighborhood $m\left(i\right)$, we can construct $\tilde{X}_j \in \reals^{N \times p_j}$. The matrix $\tilde{X}_j$ can be rescaled and then combined into a widened neighborhood matrix $\tilde{X} \in \reals^{N \times \sum_{j=1}^{J}{p_j}}$.
This neighborhood matrix $\tilde{X}$ is input to a dimensionality reduction method to recover a set of embeddings. Our final set of microenvironments is found by clustering these embeddings. Below, we apply $K$-means to the set of neighborhood-level embeddings.

\subsection{Example}

We next discuss a specific instantiation of this general procedure, describing the neighborhood and featurization choices used in Section \ref{sec:data} and implemented in NBFvis. There, $N$ gives the number of cells in one tissue section, $D$ is the number of proteins measured, and $s$ are the centers of the segmented cells.
We apply PCA to the expression matrix $X \in \reals^{N \times D}$ and then derive the reduced expression matrix $\hat{X} \in \reals^{N \times P}$.
Neighborhoods are constructed by keeping the $K$ nearest neighbors that are also within a given radius.

We use two types featurization functions $T_j$ -- quantile features and network features. For the $i^{th}$ cell's neighborhood, $Z$ quantiles $\left(q_1^{i,k},q_2^{i,k},\dots ,q_Z^{i,k}\right)$ are calculated for the $k^{th}$ protein, where $k = 1,2,\dots,P$. For the neighborhood of the $i^{th}$ cell, we derive a $PZ$-dimensional vector $\left(q_1^{i,1},q_2^{i,1},\dots,q_{Z}^{i, P}, q_{Z-1}^{i,P}\right)$. Thus, $T_{\text{quantile}}(X) : \reals^{N\times P} \to \reals^{N\times PZ}$. After featurization, we obtain an $N\times PZ$ matrix, which we call the ``quantile matrix.''
Next, consider construction of a network features. Let $G_{i}$ denote the geometric graph associated with $m_{i}$, using the metric induced by $s$. Based on $G_{i}$, we can calculate a variety of node or edge features. The associated network featurization here is $T_{\text{network}}\left(X\right) :\reals^{N\times P} \to \reals^{N\times M}$, where $M$ is the number of network statistics. For example, in the experiments below, we use the number of edges $\text{degree}\left(G_{i}\right)$ and a variety of centrality measures. We use an ensemble of 29 network-based statistics in our example, detailed in the appendix. 

The final featurization combines both quantile and network features,
\begin{align*}
    T\left(\hat{X}\right) = \left[T_{\text{quantile}}\left(\hat{X}\right), T_{\text{network}}\left(\hat{X}\right)\right].
\end{align*}
$T(X)$ is a $N \times (PZ+M)$ neighborhood matrix. Rescaling is applied to this neighborhood matrix so that every column is on a similar scale.
This rescaled neighborhood matrix is passed to UMAP to obtain low-dimensional embeddings. These embeddings can then be clustered to identify distinct microenvironments.
    
\subsection{Implementation details}

 Several subtle but important details are worth noting. Before we calculate a featurization matrix, a preliminary dimensionality reduction method is needed. First, applying dimensionality reduction decreases the computational burden of downstream analysis. Computing quantiles for each feature in a high-dimensional dataset will further increase the dimensionality. For example, computing 10 quantiles for each of 100 variables results in 1000 columns. This significantly increases the computational burden of embedding. Second, a statistical reason for dimensionality reduction is to reduce the noise in the original high-dimensional dataset. If the original data are effectively low rank, then dimensionality reduction method will reduce unnecessary noise while preserving most statistical information, which is beneficial for the following embedding.

Another detail is the rescaling of the neighborhood matrix. Although the neighborhood matrix could have hundreds or even thousands of columns, there is no need to apply a preliminary dimensionality reduction to it, since all values are approximately comparable. However, it is necessary to rescale the neighborhood matrix because the ranges of different statistics vary dramatically, causing one or two variables with large variance to dominate the whole UMAP embedding. For instance, the entries in the quantile matrix are between -1.5 and 1.5 in the TNBC dataset, but for the network matrix, it is common to have some network statistics larger than 10. These network statistics would dominate the UMAP embedding if no rescaling is applied.

\section{Visualization Design}
\label{sec:visualization}

We devise an interactive Shiny app \citep{chang2015package} to analyze outputs from the neighborhood-based analysis, supporting visualization of microenvironment differences. In this section, we discuss the design and visual queries supported by the interface.

\begin{figure}
    \centering
    \includegraphics[width=1\textwidth]{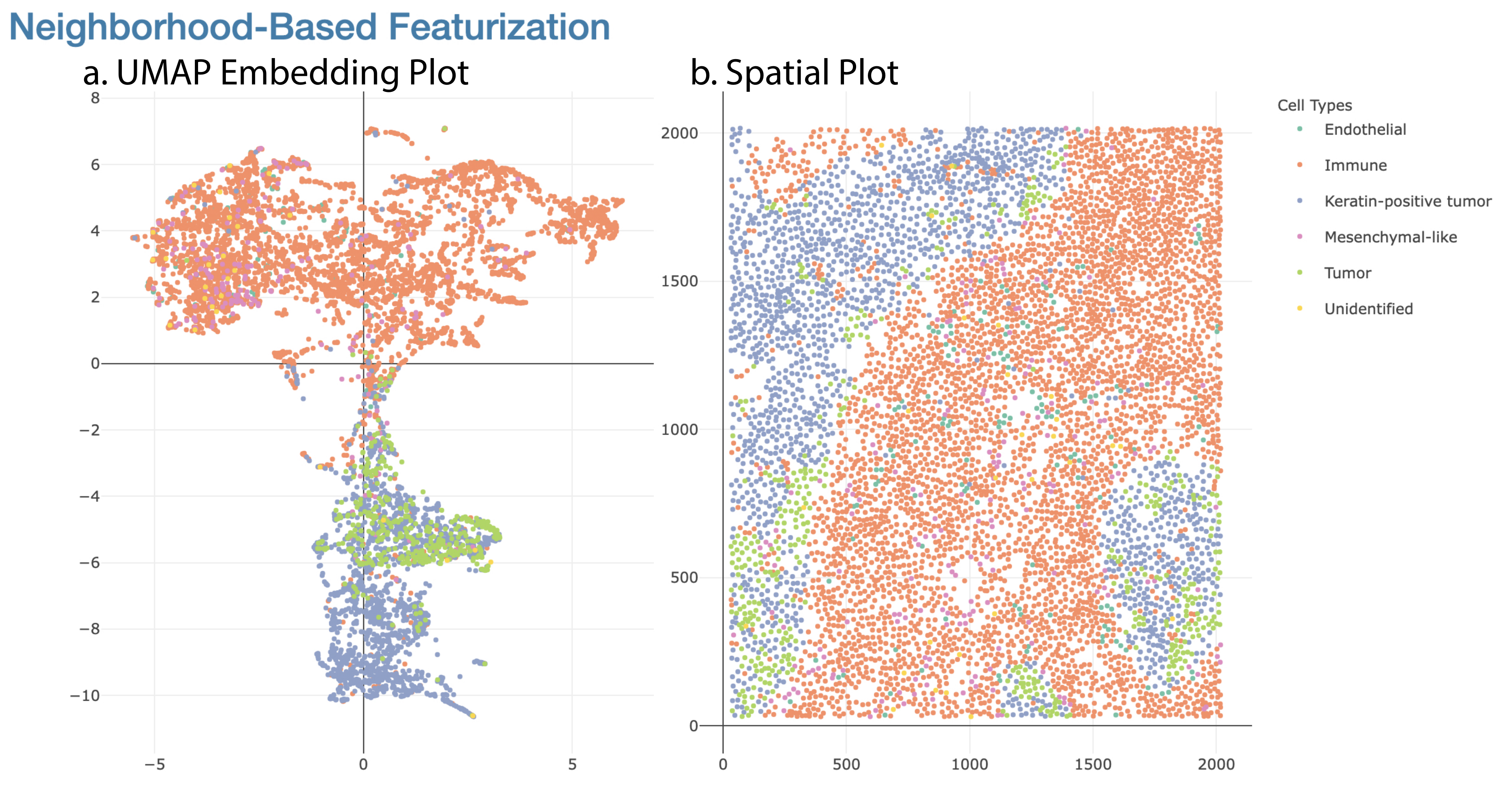}    
    \caption{The first component: the UMAP embedding and spatial plots. Part (a) is the two-dimensional embedding of the neighborhood matrix, and (b) is the original spatial layout of cell types.}
    \label{fig:shiny UMAP and spatial}
\end{figure}

Figure \ref{fig:shiny UMAP and spatial} shows the first component of the Shiny app, the UMAP embedding and linked spatial plot. This is used to relate the low-dimensional embeddings of each cell's neighborhood features to its overall spatial context. Figure \ref{fig:shiny UMAP and spatial}a is the two-dimensional UMAP embedding plot derived from the neighborhood matrix. Each point corresponds to one cell. The closer these points are, the more similar their neighborhood featurizations. To clearly visualize the distribution of cell types, the points in Figure \ref{fig:shiny UMAP and spatial}a are colored according to cell types. 
\begin{figure}
    \centering
    \includegraphics[width=1\textwidth]{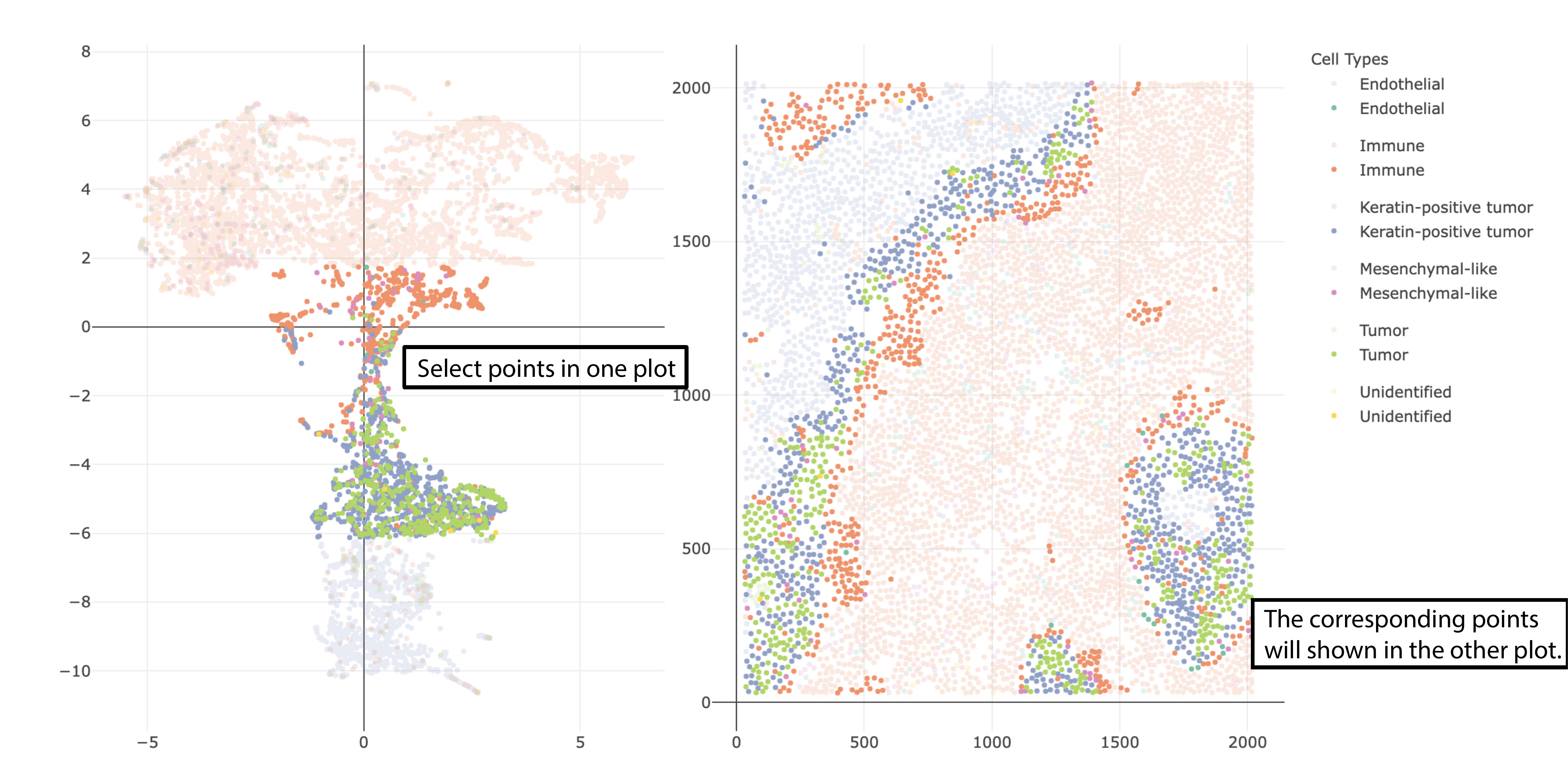}  
    \caption{Area selection}
    \label{fig:shiny area selection}
\end{figure}

\begin{figure}
    \centering
    \includegraphics[width=1\textwidth]{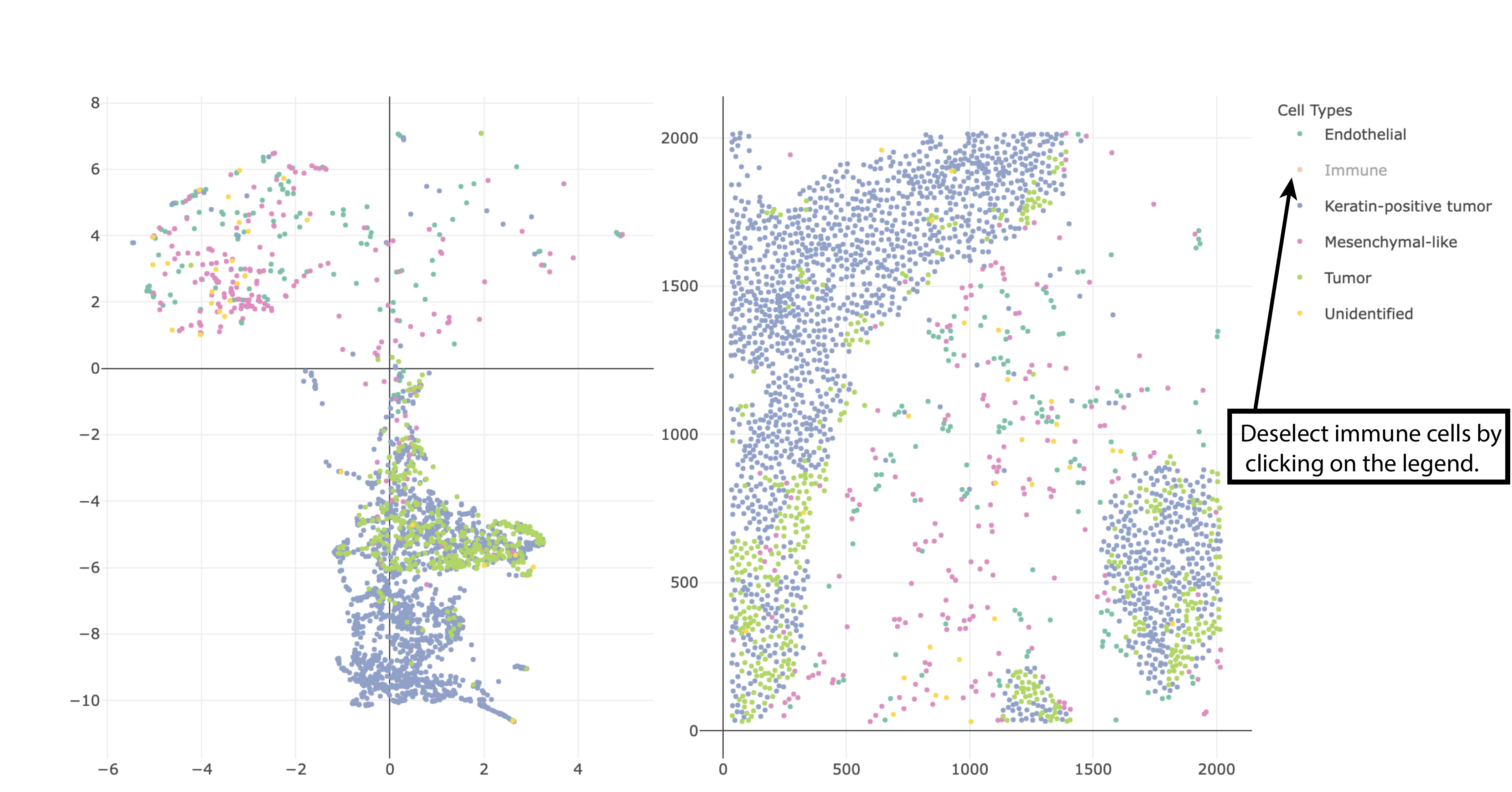}  
    \caption{Cell types can be deselected by clicking on the legend.}
    \label{fig:deselect}
\end{figure}

Figure \ref{fig:shiny UMAP and spatial}b is the spatial plot. Each point here represents a cell center, derived from the original cell polygon in the tissue section. As before, different cell types are distinguished by colors.
Furthermore, the two panels in Figure \ref{fig:shiny UMAP and spatial} are dynamically linked. When points are selected in one plot through a mouse brush, the corresponding points will also be highlighted in the other plot. Figure \ref{fig:shiny area selection} shows the highlighted points in these two plots after one such selection. 
We can click on the legend on the sidebar to deselect the these cell types so that they do not appear. Figure \ref{fig:deselect} shows the embedding plot and scatterplot after deselecting the immune cells.

\begin{figure}
    \centering
    \includegraphics[width=1\textwidth]{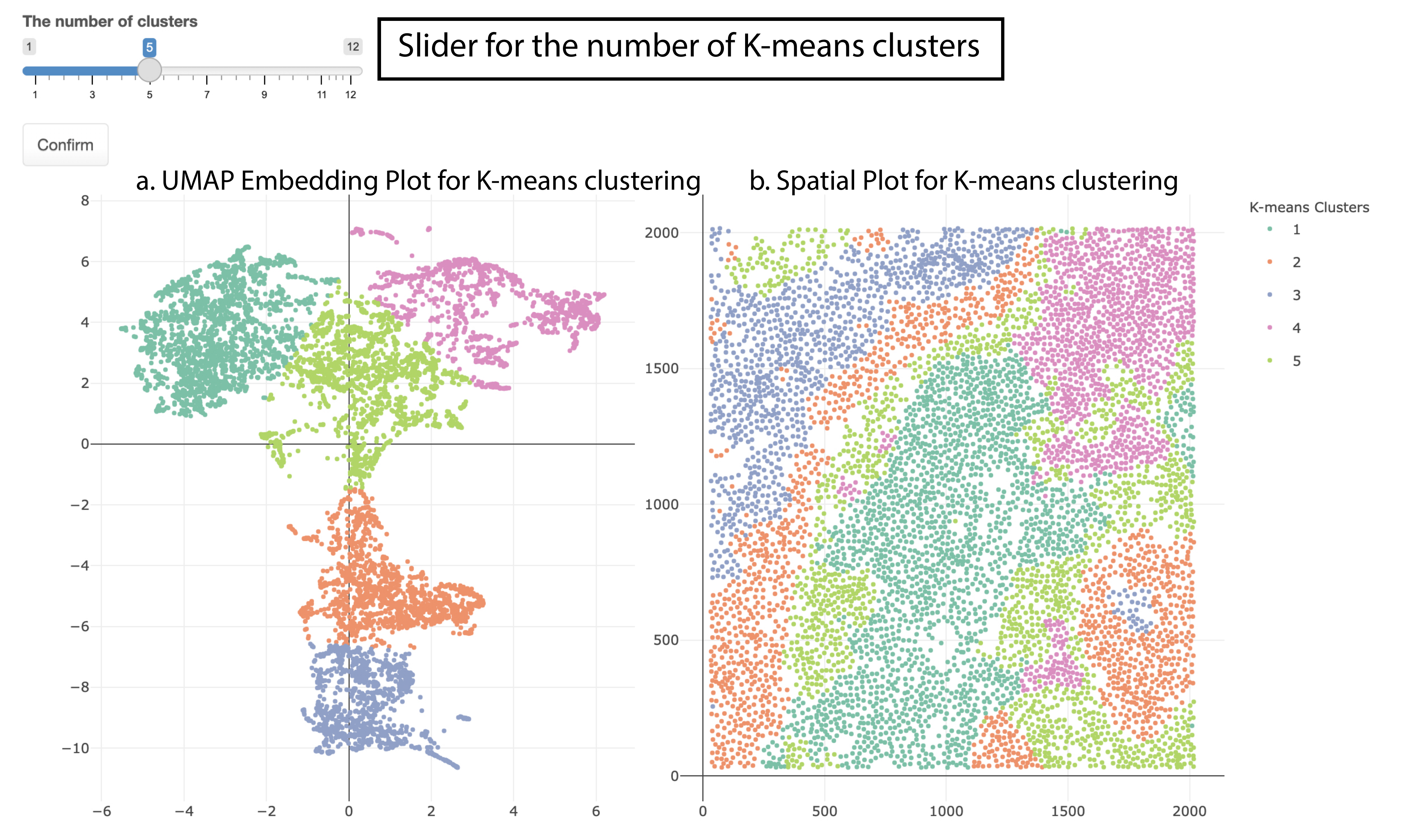}
    \caption{The second component: UMAP embedding plot and spatial plot of $K$-means clustering.}
    \label{fig:shiny $K$-means}
\end{figure}

The second component of the Shiny app shows the same embeddings, but colored by $K$-means cluster rather than cell type. For example, in Figure \ref{fig:shiny $K$-means}, the positions of points are still the same as in Figure \ref{fig:shiny UMAP and spatial}, but they are clustered into five $K$-means clusters. A slider is provided at the top of the second component in Figure \ref{fig:shiny $K$-means}, which is used for changing the value of $K$ in the $K$-means clustering.

\begin{figure}
    \centering
    \includegraphics[width=1\textwidth]{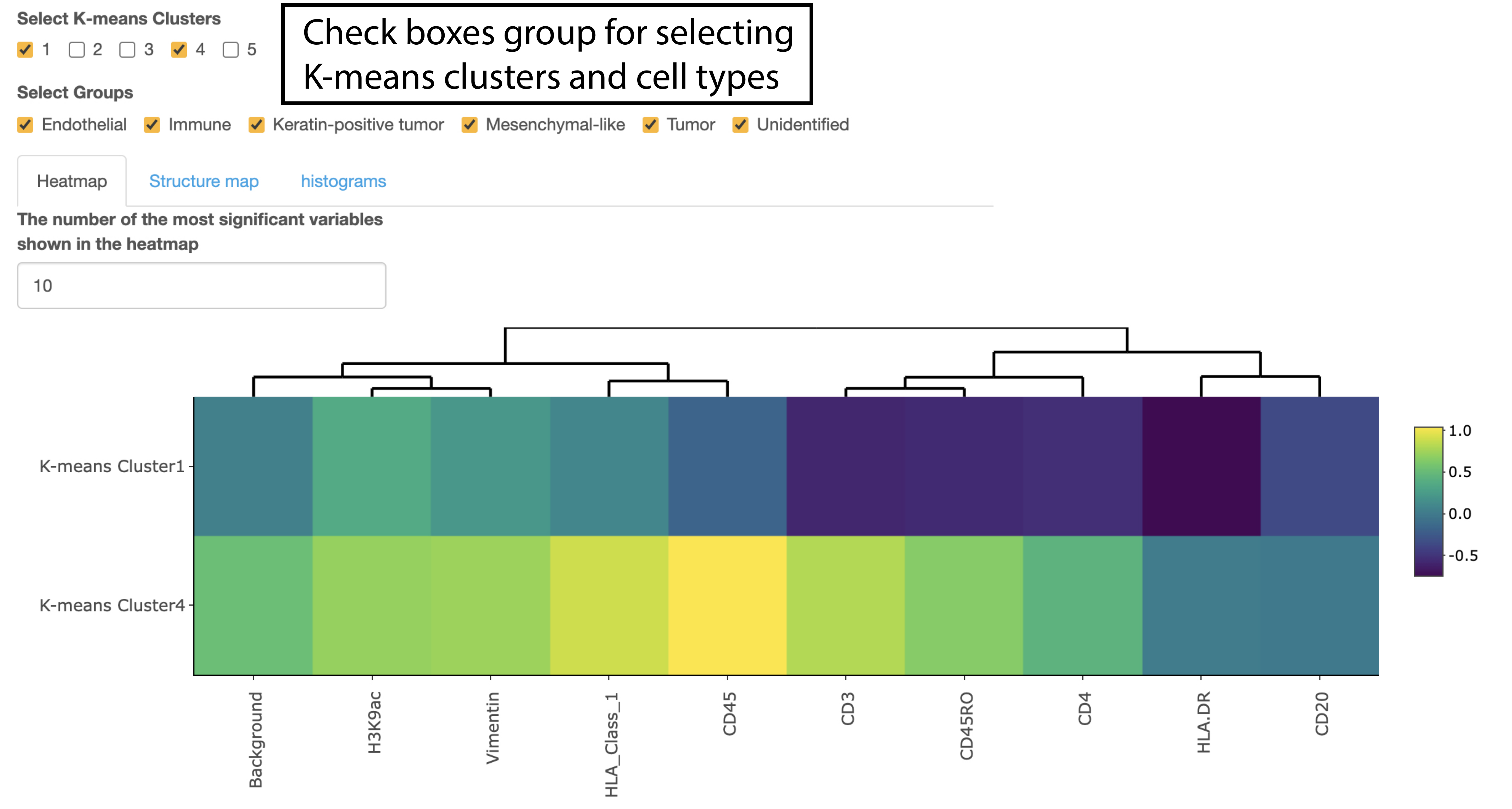}  
    \caption{The third part: Heatmap, structure plot, and histogram. These views help describe clusters identified by $K$-means.}
    \label{fig:shiny heat map}
\end{figure}

\begin{figure}
    \centering
    \includegraphics[width=1\textwidth]{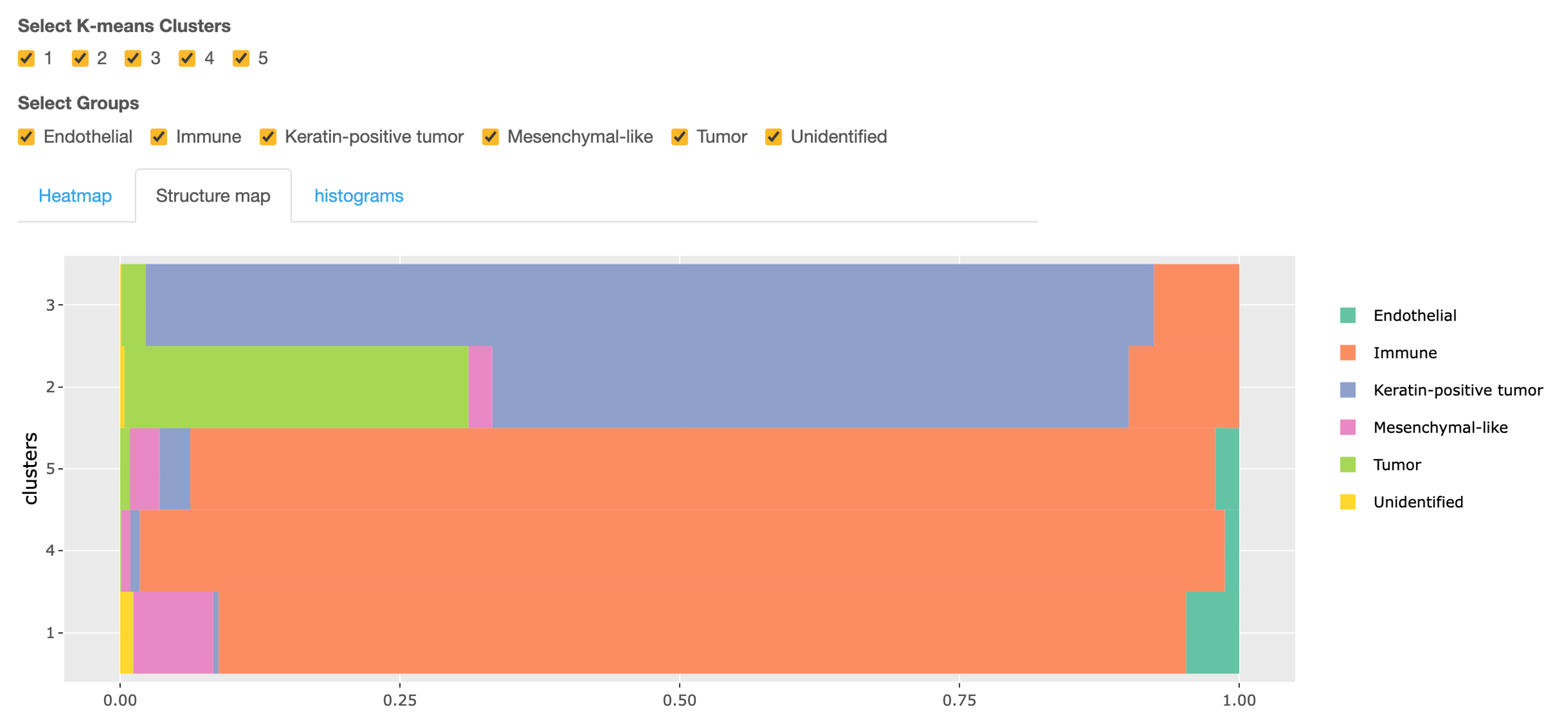} 
    \caption{Structure plot of $K$-means clusters. The dominant cell types in each clusters are shown clearly.}
    \label{fig:shiny structure plot}
\end{figure}

\begin{figure}
    \centering
    \includegraphics[width=1\textwidth]{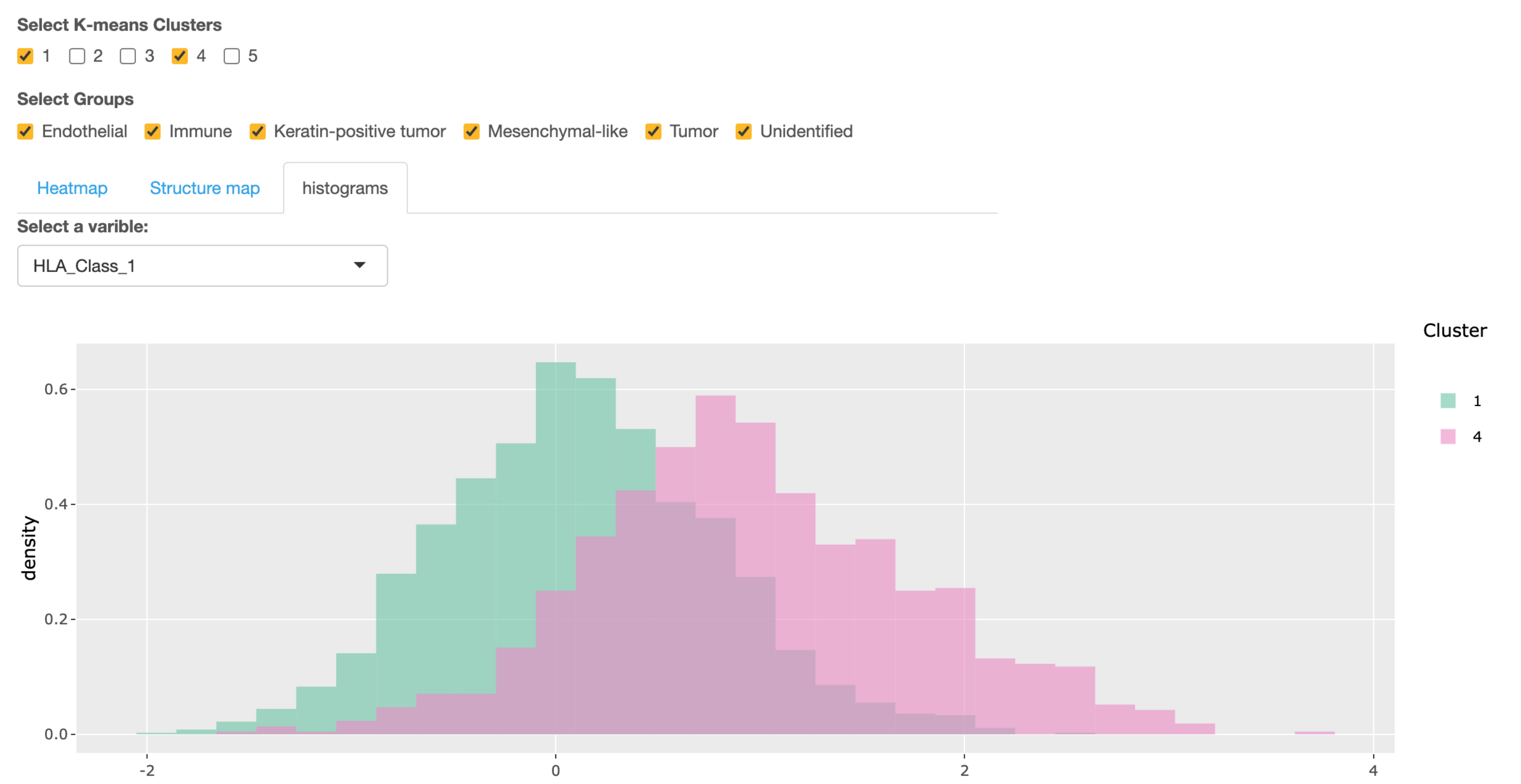}
    \caption{Histogram of expressions among $K$-means clusters. Users could select different expressions by the selection input box above the histogram.}
    \label{fig:shiny histogram}
\end{figure}

The third component of this Shiny app supports the comparison of expression levels across $K$-means clusters using a heatmap, structure plot, and histogram; see Figure \ref{fig:shiny heat map}. There are three tab panels with which we can switch between these three plots.
Before making further comparison, we can filter to cells of interest using the check boxes at the top of Figure \ref{fig:shiny heat map}. Two groups of check boxes are offered to select the cell types and $K$-means clusters to focus on.
Based on the filtered cells, a expression heatmap of $K$-means clusters is provided in Figure \ref{fig:shiny heat map}. By default, it shows the top 10 most differentially expressed features across the selected clusters, based on the median of expression value in each cluster. A numeric input is offered above the heatmap -- this controls the number of features appearing in the heatmap.
The structure plot of the selected $K$-means clusters is provided in Figure \ref{fig:shiny structure plot}, with which we can see the proportion of each cell type across every cluster.
The histogram of expression is available to compare the selected feature's expression across clusters. For example, Figure \ref{fig:shiny histogram} is the histogram of the HLA Class 1 content in Cluster 1 and 4. Note that a selection input box is offered above the histogram to change the selected feature easily.

\begin{figure}
    \centering
    \includegraphics[width=1\textwidth]{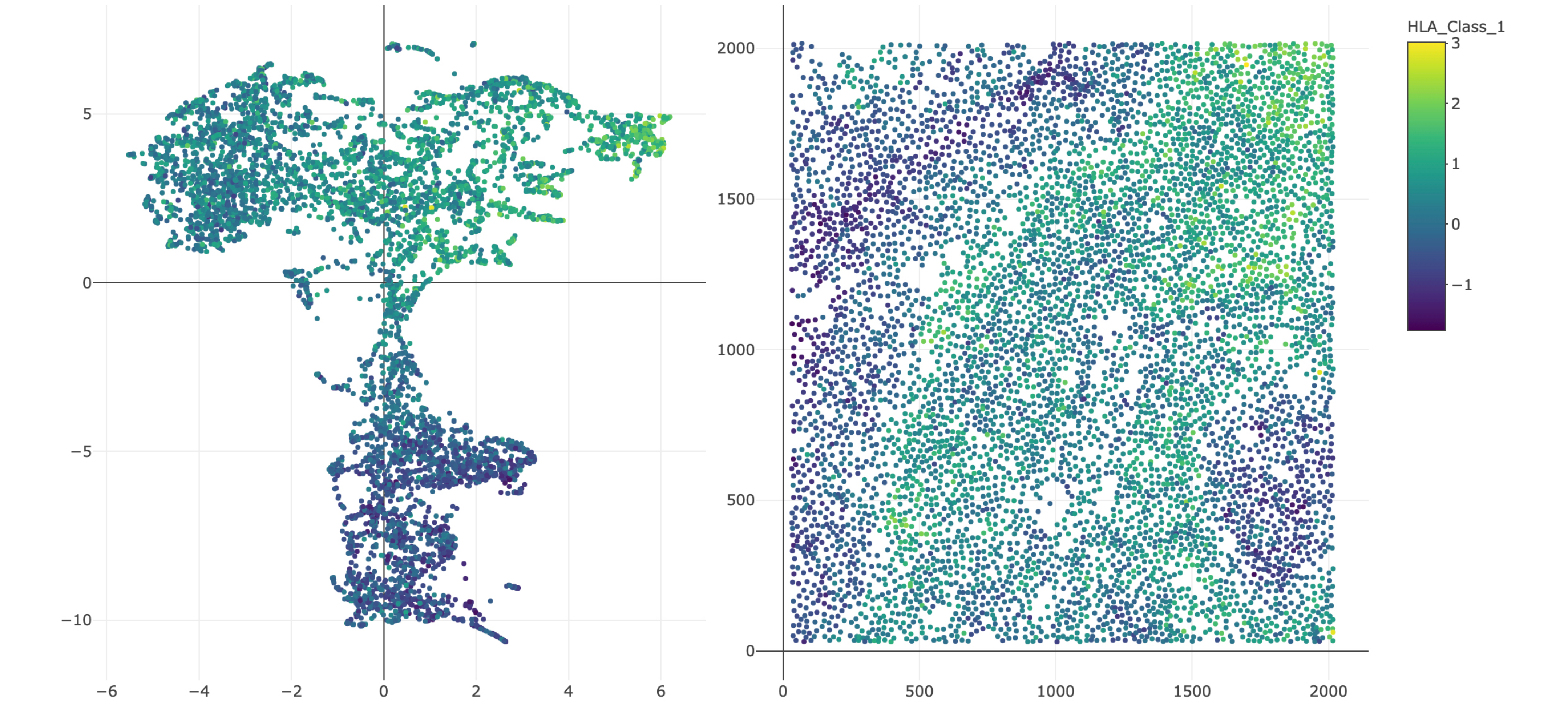}
    \caption{Expression plot of selected cells. Instead of coloring by cell type or $K$-means cluster, each cell is shaded according to the selected genomic feature.}
    \label{fig:shiny expression plot}
\end{figure}

To show the spatial distribution of a specific feature's expression, another combination of the embedding and spatial plot is provided in Figure \ref{fig:shiny expression plot}. The colors of the points in Figure \ref{fig:shiny expression plot} are reflect HLA Class 1 content. This expression plot highlights spatial characteristics of the expression content. In this case, expression is elevated in immune cells, especially those closest to the tumor-immune boundary.    

\section{Data analysis}
\label{sec:data}

To illustrate our approach and package, we re-analyze the Triple Negative Breast Cancer (TNBC) dataset of \cite{keren2019mibi}. To study this data, \cite{chen2020modeling} proposed Spatial-LDA, which was found to reveal novel microenvironments. Spatial-LDA models the distribution of cell types within neighborhoods but does not model protein expression directly. In contrast, our proposal considers quantitative protein measurements and network statistics within spatial neighborhoods. Here, we choose the tissue section of Patient 4, which has 6643 cells consist of six cell types,  immune cells (62.6\%), keratin-positive tumor cells (25.2\%), tumor cells (6.4\%), mesenchymal-like cells (3.2\%), endothelial cells (1.9\%), and unidentified cells (0.5\%). We use 41 expression variables, two-dimensional coordinates of cell centers, and cell types for further analysis.

The first step is to construct the neighborhood quantile matrix. We apply PCA to reduce the dimension of the expression matrix. We keep 19 principal components, which is the smallest number of components required to explain 90\% of the variance. These components are labelled as $PC_1,...,PC_{19}$. Next, neighborhoods are defined using a radius of 60 pixels. We only include the cells among the top 40 nearest neighbors to the center cell of the neighborhood. Quantiles for each principal component are calculated based on neighborhoods. To avoid the influence of extreme values, only quantiles $q_{0.10}, q_{0.15},...,q_{0.90}$ are included. Hence, we derive a $6643\times323$ quantile matrix of neighborhoods after featurization. 
The second step is to obtain the network matrix of the neighborhoods. We again use a radius of 60 pixels to define neighborhoods and keep only the 40 closest cells. Networks are constructed based on these neighborhoods. We link cells whose centers are within 30 pixels of one another. Then, 29 network statistics are calculated according the neighborhood networks; most of these network statistics are different kinds of network centralities. This results in a $6643\times29$ neighborhood network matrix.

The third step is to combine the quantile and network matrices together into an extended neighborhood matrix. The network matrix is rescaled in this step. The result is a $6643\times352$ neighborhood matrix. The final step applies dimensionality reduction and clustering to the neighborhood matrix. We apply UMAP to the neighborhood matrix to generate 2-dimensional embeddings of each cell. $K$-means is applied to the UMAP embeddings to find potential clusters. These can be interpreted as microenvironments. 

\begin{figure}
    \centering
    \includegraphics[width=1\textwidth]{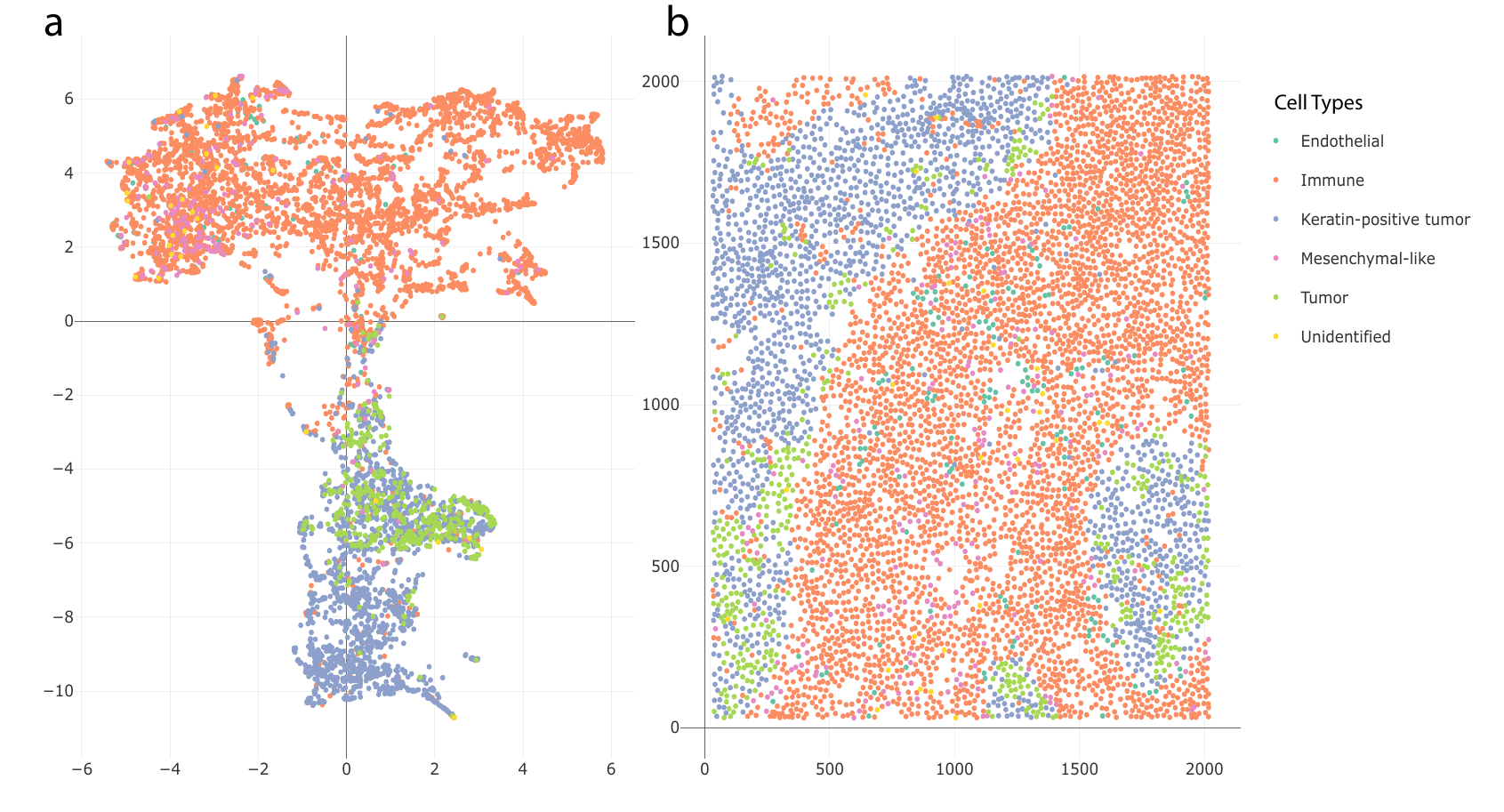}
    \caption{UMAP embedding and spatial plots using neighborhood-based featurization. Panel (a) is the UMAP embedding plot colored in cell types. Panel (b) is the spatial plot of the real positions of cell centers. We observe a transition zone between clusters of tumor cells and immune cells in part (a).}
    \label{fig:UMAP embedding and spatial plot}
\end{figure}

\begin{figure}
    \centering
    \includegraphics[width=1\textwidth]{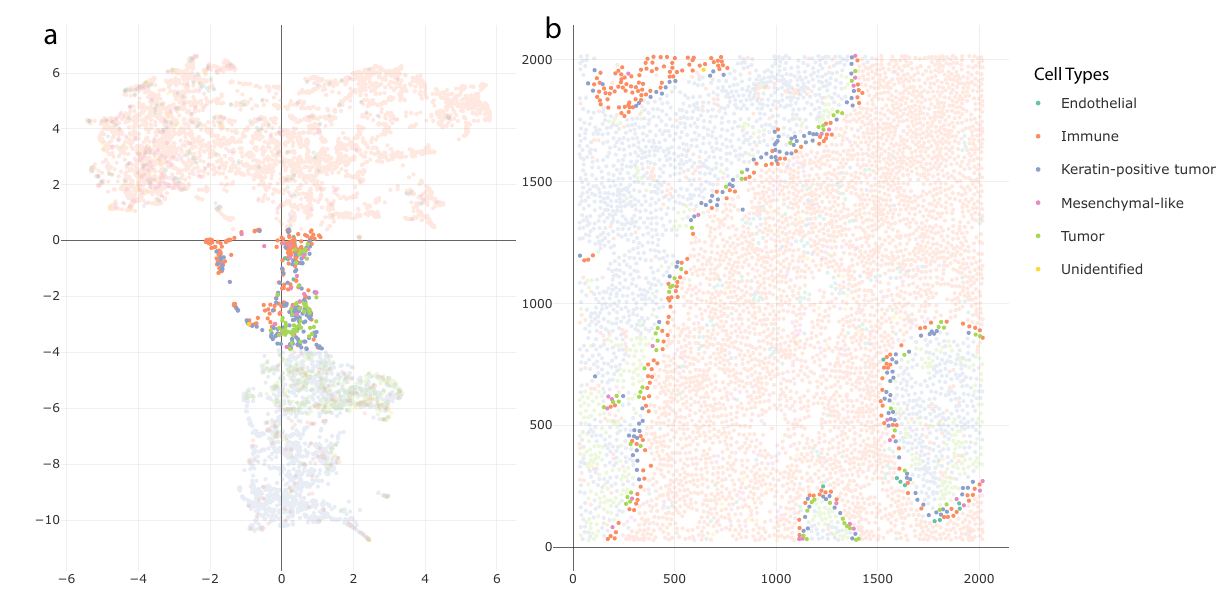}
    \caption{Transition zone in the UMAP embedding and spatial plots. The corresponding cells whose embeddings are in the transition zone in panel (a) are located close to the tumor-immune boundary in panel (b).}
    \label{fig:transition zone}
\end{figure}

We use a Shiny app implemented in NBFvis to explore the resulting of UMAP embeddings and $K$-means clusters. 
Figure \ref{fig:UMAP embedding and spatial plot} shows the UMAP embeddings and spatial plot of the neighborhood matrix. Figure \ref{fig:UMAP embedding and spatial plot}a gives the embeddings based on the reduction of the neighborhood matrix. The points in the embedding plot are colored according to their cell types. There are two main clusters in the embedding plot, composed primarily of immune and tumor cells, respectively. These two clusters are connected by a transition zone of a mixture of tumor and immune cells. Figure \ref{fig:UMAP embedding and spatial plot}b is the spatial plot of the cells in the tissue section. By selecting the transition zone in the embedding plot, we find that the cells in this area are located on the boundary of immune cells, tumor cells, and keratin-positive tumor cells. This is shown in Figure \ref{fig:transition zone}.

\begin{figure}
    \centering
    \includegraphics[width=1\textwidth]{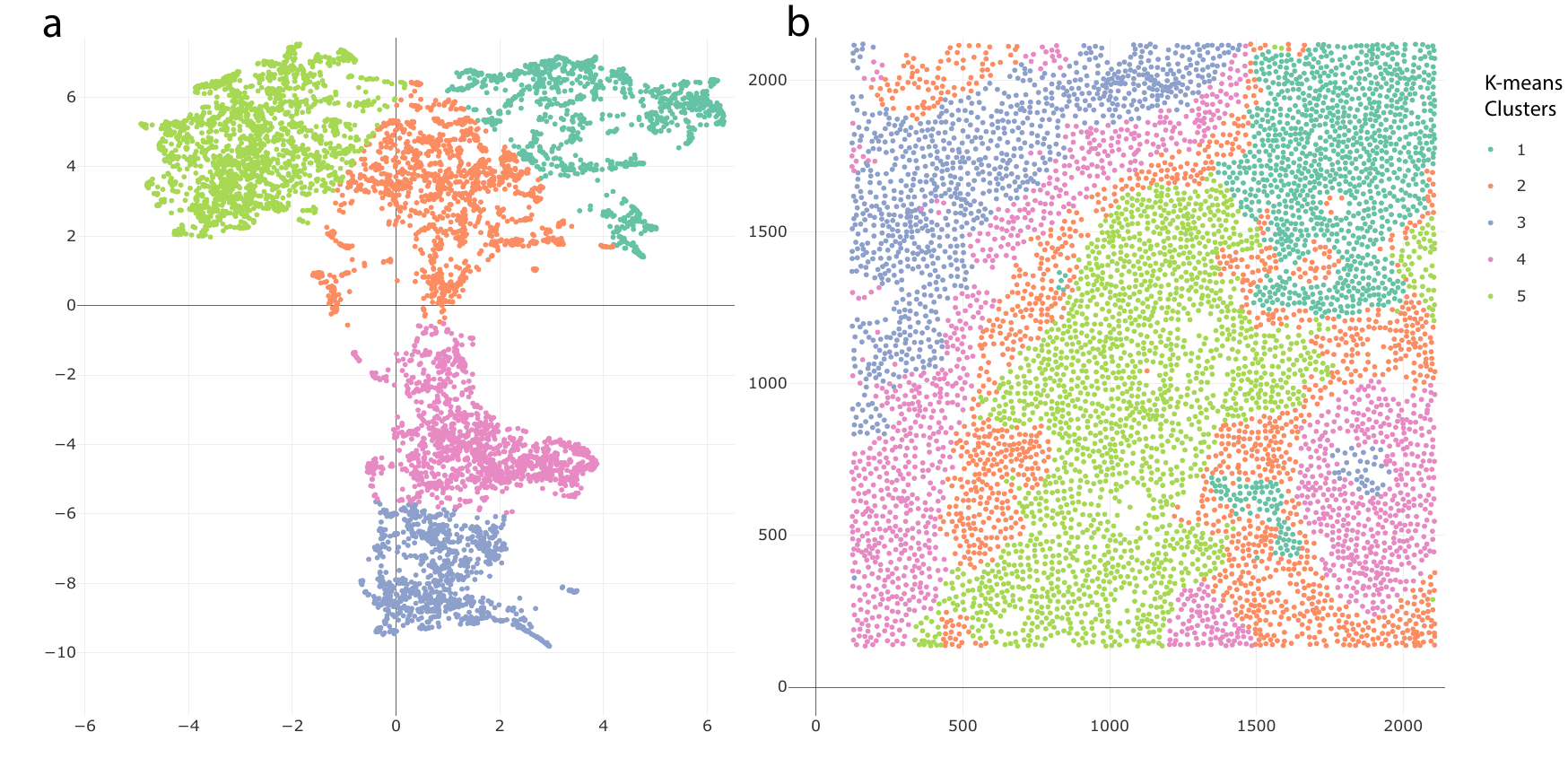}
    \caption{$K$-means embedding and spatial plots with $K = 5$. Clusters in panel (b) are spatially consistent. There are two special clusters on the tumor-immune boundary, whose embeddings are in the transition zone in panel (a).}
    \label{fig:K-means embedding and spatial plot}
\end{figure}

$K$-means clustering applied to the UMAP embeddings suggests potential microenvironments. Figure \ref{fig:K-means embedding and spatial plot} shows clustering results with $K= 5$. The clusters are distinguished by their colors. In the embedding plot Figure \ref{fig:K-means embedding and spatial plot}a, the embeddings are divided into 5 clusters, and in the spatial plot Figure \ref{fig:K-means embedding and spatial plot}b, the corresponding locations of these clusters are shown. One finding is that the clusters in the embedding space are spatially consistent. 

 In the Figure \ref{fig:K-means embedding and spatial plot}b, two microenvironments are founded among the tumor cells and keratin-positive tumor cells, Cluster 3 in the inner part of the tumor cell groups and Cluster 4 close to the boundary of immune cells. This mirrors the findings of \cite{chen2020modeling}. Another finding is that there is a special immune cell microenvironment, Cluster 2, lying on the boundary of immune cells, tumor cells, and keratin-positive tumor cells. This microenvironment is distinguished from the immune microenvironment in the inner part of immune cell groups, which is Cluster 5 in Figure \ref{fig:K-means embedding and spatial plot}b. Notice that Clusters 4 and 5, which are the microenvironments close to the tumor-immune boundary, are in the transition zone in the UMAP embedding plot in Figure \ref{fig:transition zone}. Moreover, another microenvironment, Cluster 3, is found in the top-left corner of Figure.\ref{fig:K-means embedding and spatial plot}b, separate from the previous two immune microenvironments, Clusters 2 and 5.

\begin{figure}
    \centering
    \includegraphics[width=1\textwidth]{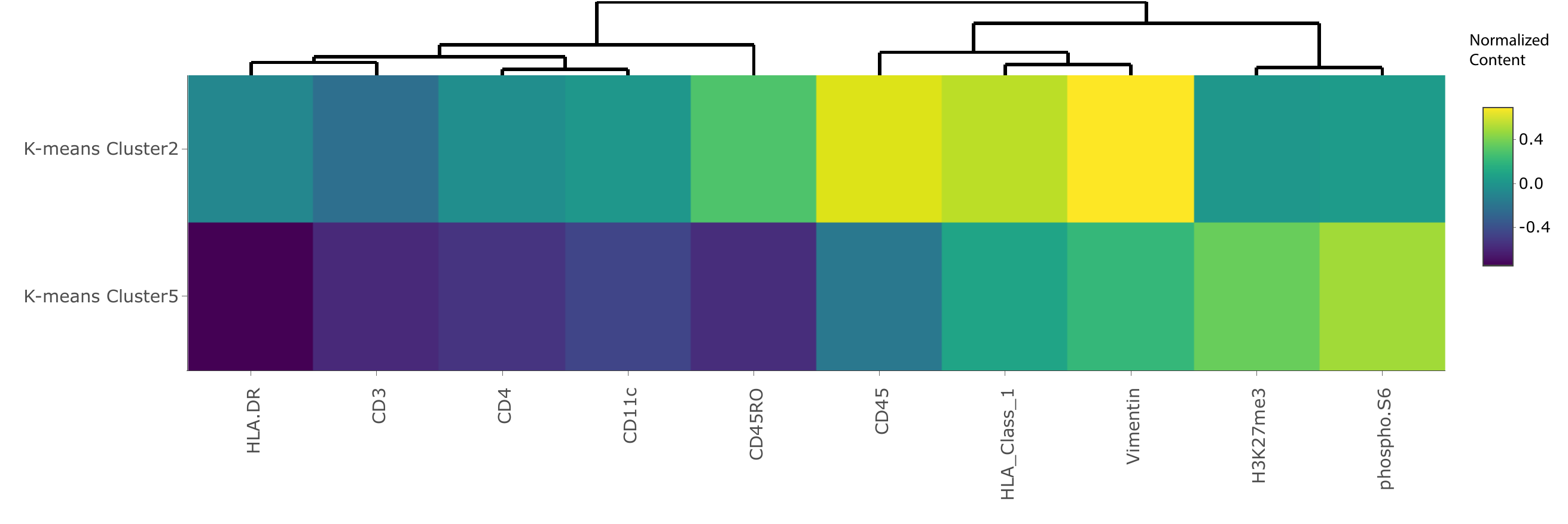}
    \caption{Heatmap of expressions in Cluster 2 and 5. Cluster 2 is on the tumor-immune boundary and Cluster 5 is in the inner part of immune cell groups. The most obvious difference in expressions between these two clusters are CD45 and CD45RO. Cluster 5 has significantly lower CD45 and CD45RO content than Cluster 2.}
    \label{fig:heat map - inner immune vs boundary immune}
\end{figure}

\begin{figure}
    \centering
    \includegraphics[width=1\textwidth]{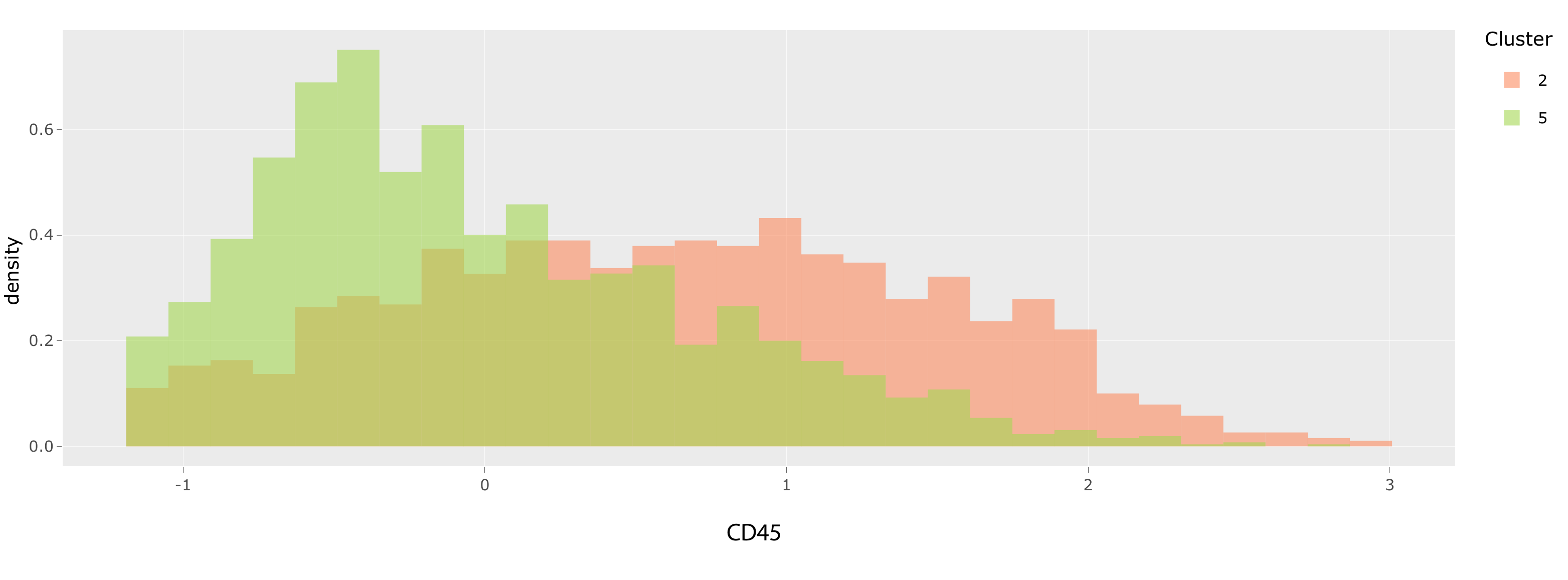}
    \caption{Histograms of CD45 across Clusters 2 and 5, highlighting elevated CD45 levels in immune cells closer to the tumor-immune boundary. Histograms for different features can be selected using the interface, and the choice can be guided by a heatmap like in Figure \ref{fig:heat map - inner immune vs boundary immune}.}
    \label{fig:histogram CD45 in Cluster 2 and 5}
\end{figure}

\begin{figure}
    \centering
    \includegraphics[width=1\textwidth]{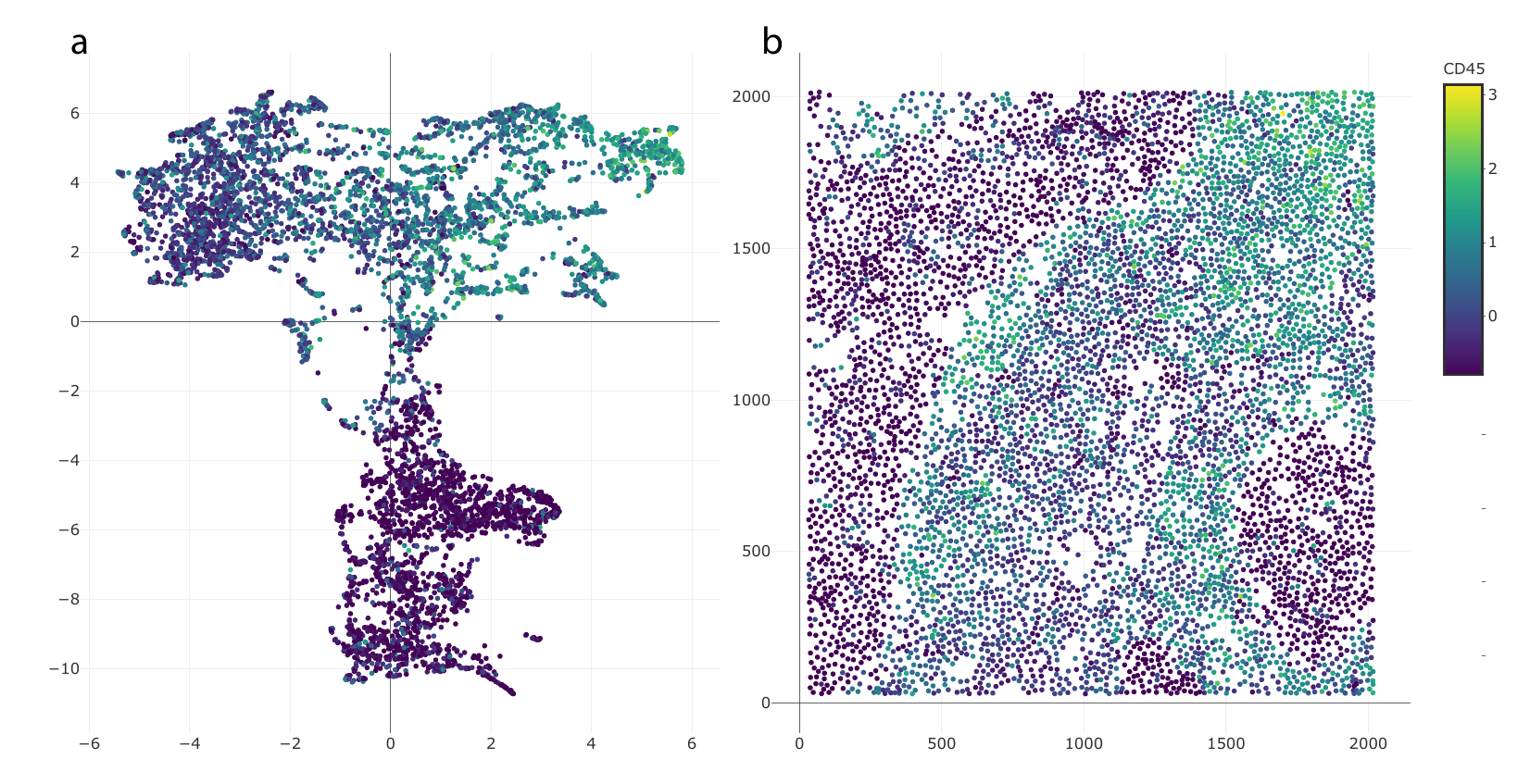}
    \caption{UMAP embedding and spatial plots shaded in according to measured CD45. The existence of brighter cells near the tumor-immune boundary is consistent with Figures \ref{fig:heat map - inner immune vs boundary immune} and \ref{fig:histogram CD45 in Cluster 2 and 5}. This view also reveals elevated CD45 in the top-right region, corresponding to Cluster 3.}
    \label{fig:CD45 in Cluster 2 and 5}
\end{figure}

Next, we explore the differences between these microenvironments by studying their expression patterns. Figure \ref{fig:heat map - inner immune vs boundary immune} is the heatmap of the inner and boundary immune microenvironments, which are Clusters 2 and 5 in the Figure \ref{fig:K-means embedding and spatial plot}b, respectively. The heatmap shows the top 10 most differentially expressed proteins between these two clusters, determined by the differences between medians of expressions in each group. We choose the two most differentially expressed proteins, CD45 and CD45RO, for further exploration. The histograms in Figure \ref{fig:histogram CD45 in Cluster 2 and 5} show the contents of CD45 across these two microenvironments. The inner immune microenvironment has a right-skewed distribution of CD45, indicating that many cells in this microenvironment have low content of CD45. In contrast, the distribution of CD45 in the boundary immune microenvironment is significantly higher than that in the inner immune microenvironment. Figure \ref{fig:CD45 in Cluster 2 and 5} is the expression plot of CD45, this confirms that cells along the tumor-immune boundary have elevated CD45.

\begin{figure}
    \centering
    \includegraphics[width=1\textwidth]{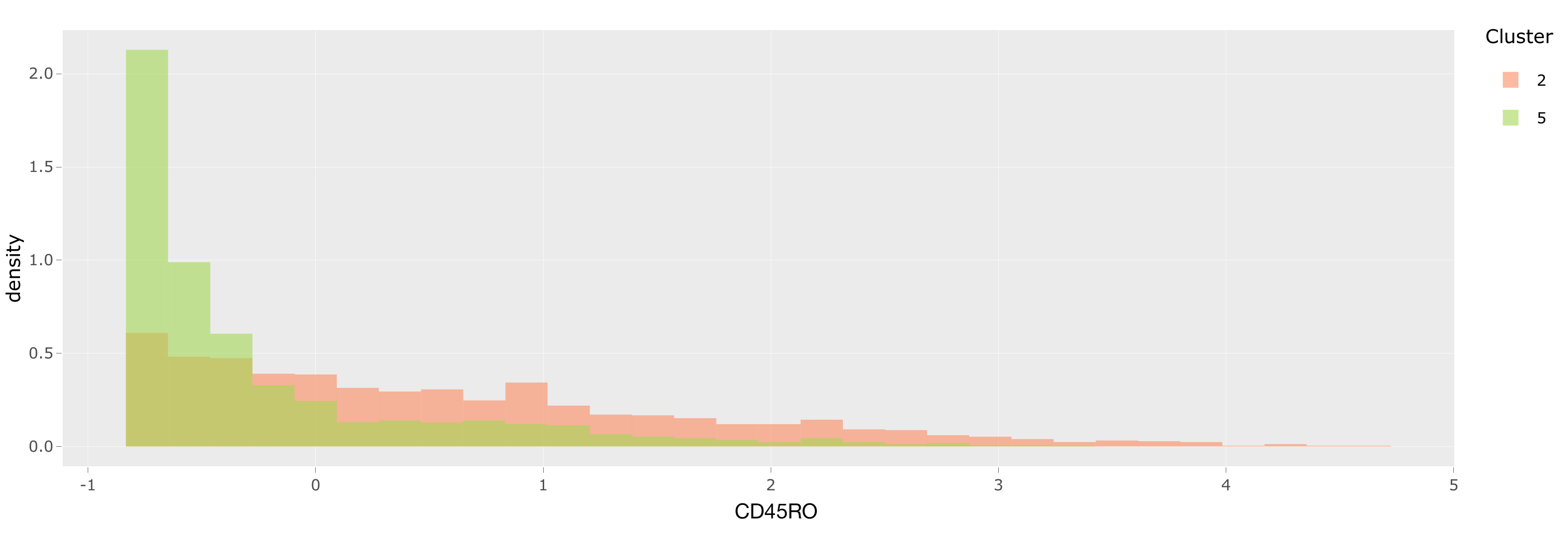}
    \caption{The analog of Figure \ref{fig:histogram CD45 in Cluster 2 and 5} for CD45RO, another marker found to be differentially expressed across Clusters 2 and 5. In contrast to CD45, the distribution in both clusters is strongly right-skewed, even after the preprocessing applied by \cite{keren2019mibi}. }
    \label{fig:histogram CD45RO in Cluster 2 and 5}
\end{figure}

\begin{figure}
    \centering
    \includegraphics[width=1\textwidth]{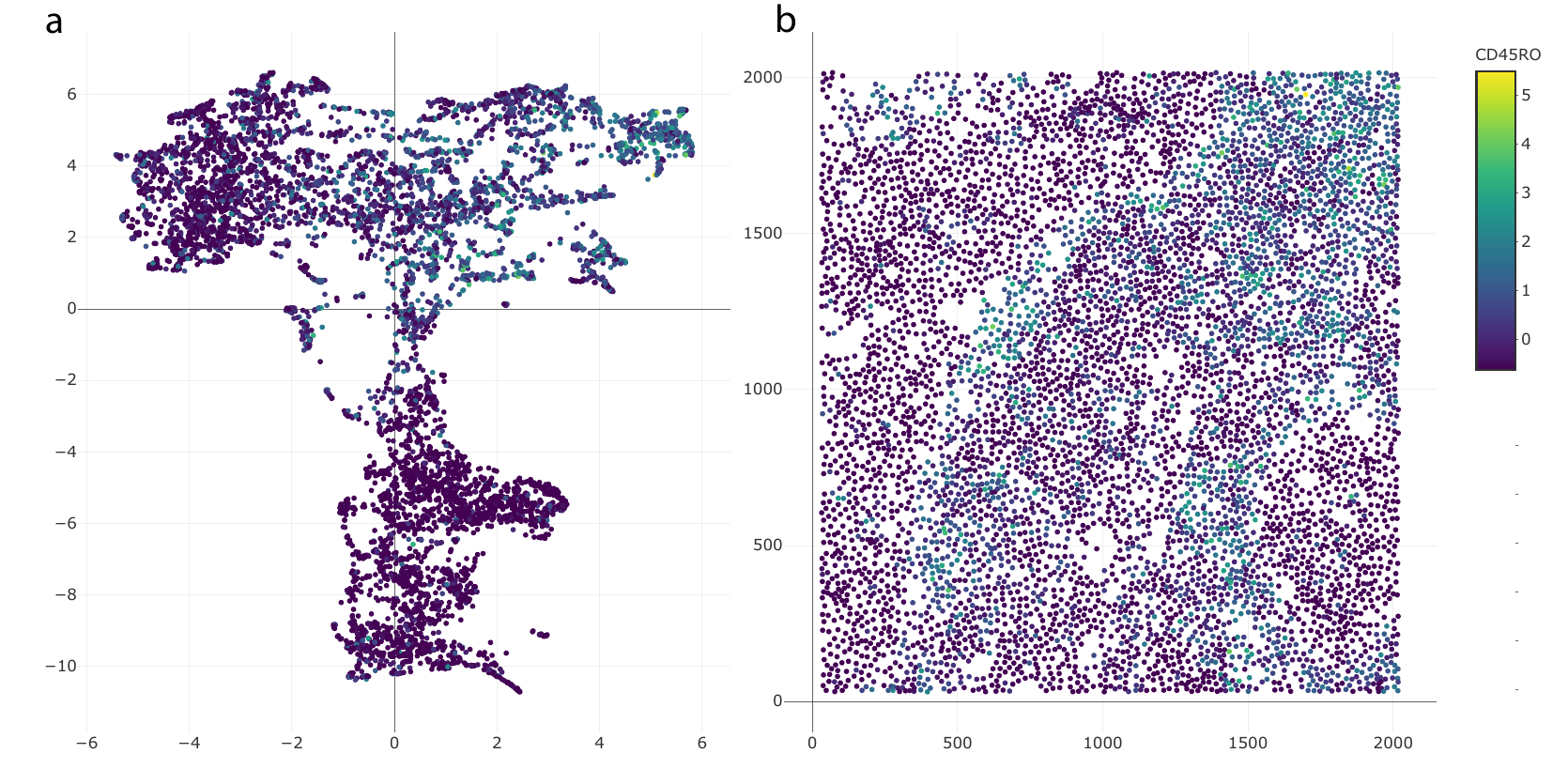}
    \caption{The analog of Figure \ref{fig:CD45 in Cluster 2 and 5} for CD45RO. This marker's spatial expression structure is similar to that for CD45. The fact that more cells are shaded darkly reflects the right skew observed in the histograms in Figure \ref{fig:histogram CD45RO in Cluster 2 and 5}.}
    \label{fig:CD45RO in Cluster 2 and 5}
\end{figure}

Checking the histogram and spatial expression of CD45RO in the inner and boundary immune microenvironments, we arrive at similar conclusions. Figure \ref{fig:histogram CD45RO in Cluster 2 and 5} is the histogram of these two microenvironments. The histogram for the inner immune microenvironments has a peak near the minimal value, which does not appear on the histogram of the boundary immune microenvironments. It shows that there are lower contents of CD45RO in the inner immune microenvironment but higher contents of CD45R0. Figure\ref{fig:CD45RO in Cluster 2 and 5} also shows that there is a lighter boundary on the tumor-immune cells, highlighting this microenvironment.

\subsection{Cell-level approach}

We also use the visualization tool to show the UMAP embedding and clustering results when directly applied to the original cell-level protein expression matrix. We use the same preprocessing as \cite{keren2019mibi}. This serves as a reference point against which to compare the proposed neighborhood-based featurization.

\begin{figure}
    \centering
    \includegraphics[width=1\textwidth]{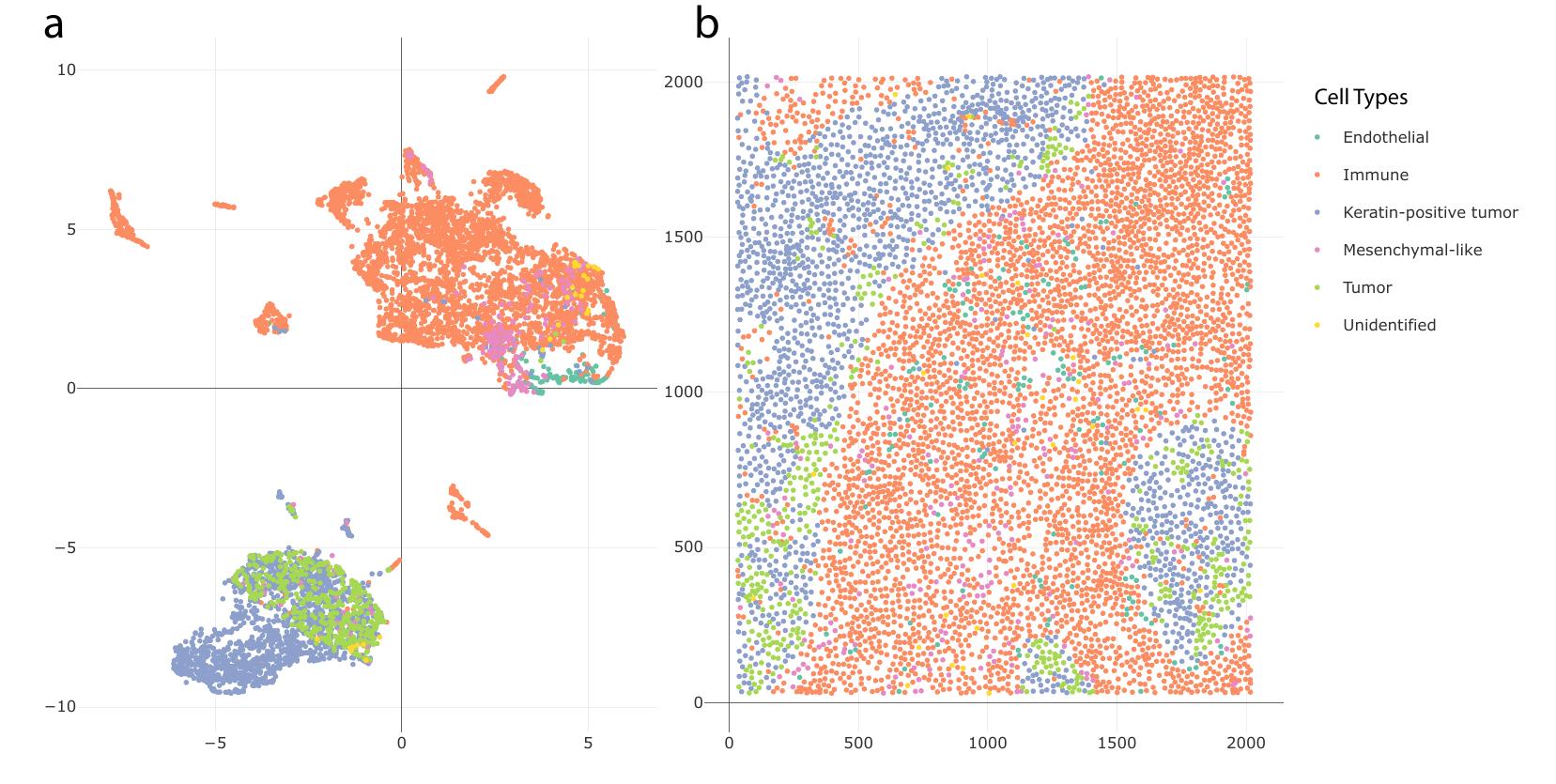}
    \caption{The UMAP embedding and spatial plots obtained without neighborhood features. Cells are shaded by cell type. Compare with Figure \ref{fig:UMAP embedding and spatial plot}.}
    \label{fig:traditional UMAP embedding and spatial plots}
\end{figure}

Figure \ref{fig:traditional UMAP embedding and spatial plots} gives the UMAP embedding and spatial plot by the cell-level approach. We find two clusters in the Figure \ref{fig:traditional UMAP embedding and spatial plots}a, one mainly made up of immune and tumor cells, respectively. The result is similar to the simulation, where UMAP embeddings are dominated by the differences between cell types and microenvironments are hardly distinguished. 

\begin{figure}
    \centering
    \includegraphics[width=1\textwidth]{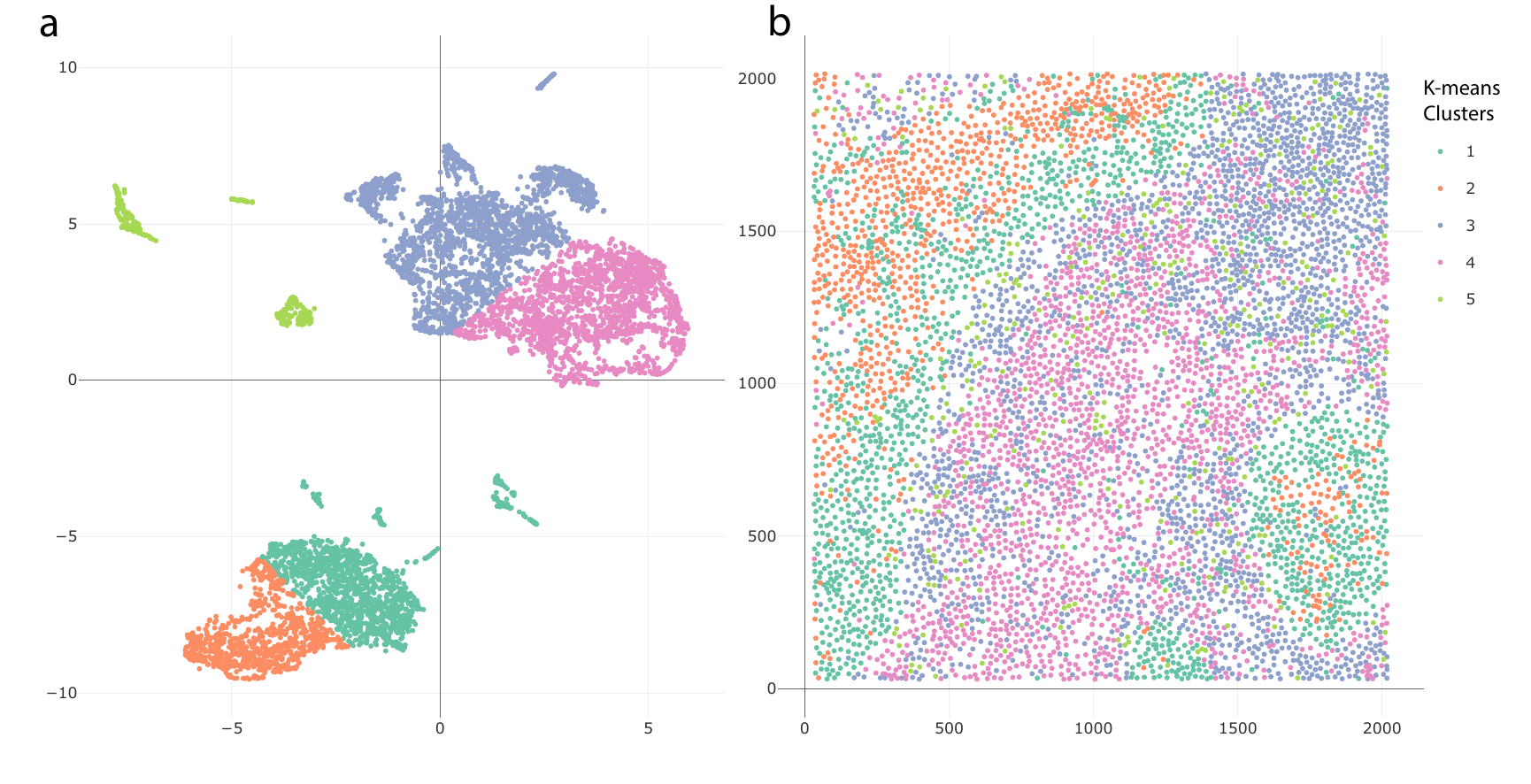}
    \caption{A version of Figure \ref{fig:traditional UMAP embedding and spatial plots} where cells are shaded by $K$-means clusters found in the embedding on the left. Sub-cell type variation in the embedding plot does not correspond to spatially meaningful microenvironments. Compare with Figure \ref{fig:K-means embedding and spatial plot}.}
    \label{fig:traditional K-means}
\end{figure}

\begin{figure}
    \centering
    \includegraphics[width=1\textwidth]{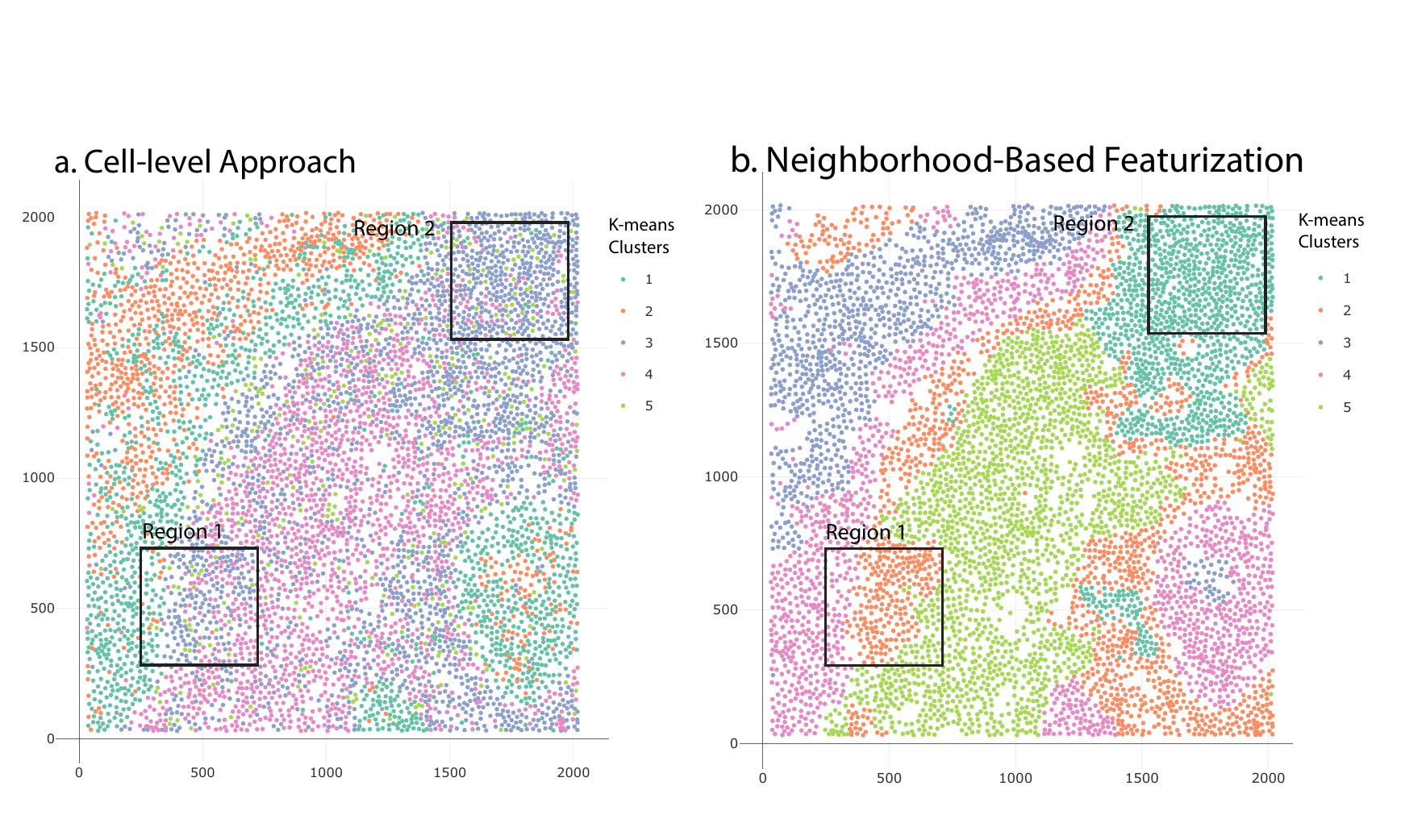}
    \caption{A direct comparison of the spatial plots from Figures \ref{fig:K-means embedding and spatial plot} and \ref{fig:traditional K-means}. Microenvironments with similar expression patterns (and stable cell type mixtures) are enclosed in black boxes. Microenvironments are more clearly visible when using neighborhood-based featurization.}
    \label{fig:comparison}
\end{figure}

Figure \ref{fig:traditional K-means} shows the clustering results after $K$-means clustering with $K = 5$. The clustered microenvironments are mixed with each other; in particular, it is difficult to distinguish a tumor-immune boundary microenvironment. 
Figure \ref{fig:comparison} compares the clustered spatial plots based on the cell-level and neighborhood-based approaches. In Figure \ref{fig:comparison}a, the cells in Region 1 are a mixture of three microenvironments derived from the cell-level approach. It is difficult to identify which microenvironment this region belongs to. Although Region 2 of Figure \ref{fig:comparison}a is mainly composed of Cluster 3,  there are cells from Clusters 4 and 5 distributed throughout. Though in principle it is possible to distinguish microenvironments based on particular mixture patterns across cell types, doing so requires much more effort than examining the neighborhood-based representation.

Compared with the cell-level approach, the neighborhood-based featurization has a noticeably clearer clustering result. In Region 1 of Figure \ref{fig:comparison}b, the cells in the boundary of tumor cells are spatially consistent according to their own cell types. 
Further, in Region 2 of Figure \ref{fig:comparison}b, we observe a dominant microenvironment without needing to parse mixed patterns of cell types.

Overall, the neighborhood-based featurization provides representations with better spatial consistency, simplifying the discovery of microenvironments.

\section{Package}
\label{sec:package}

We next summarize how to use NBFvis to implement the proposed workflow. We first load the packages and dataset we need. The dataset \texttt{patient4} is a $6643\times59$ data frame of all cells in the tissue section of Patient 4 in the TNBC data \citep{keren2019mibi}. We have added two columns named \texttt{x\_center} and \texttt{y\_center}, which are the coordinates of the calculated cell centers from the spatial raster data.

\begin{verbatim}
    library(NBFvis)
    library(dplyr)
    data(patient4)
\end{verbatim}

We select 41 variables from \texttt{dsDNA} to \texttt{HLA Class 1}, most of which are proteins and cell type markers. The \texttt{quantile\_matrix} function generates the quantile matrix from each cell's neighborhood. 

\begin{verbatim}
    Quantiles_patient4 <- quantiles_matrix(
        data = patient4 %>% select(dsDNA:HLA_Class_1),
        coordinate = patient4 %>% select(x_center,y_center),
        index = patient4$index,
        NN = 40, 
        distance = 60,
        min_percentile = 0.1,
        max_percentile = 0.9,
        quantile_number = 17,
        method = pca_)
\end{verbatim}

The function \texttt{network\_matrix} first builds the network inside the neighborhood and then calculates the corresponding network statistics using the argument given by \texttt{fun}. In this example, we use the function \texttt{centralities}, also exported by our package. 

\begin{verbatim}
    centrality_patient4 <- network_matrix(
        coordinate = patient4 %>% select(ends_with("_center")),
        index = patient4$index, 
        radius = 60,
        NN = 40,
        edge = 30,
        fun = centralities,
        length_output = 29,
        name_output = NULL)
\end{verbatim}

The scales of these two matrix are not the same, which means rescaling is needed. Here we remove \texttt{Column}, \texttt{index} and \texttt{n\_neighborhood} in the quantile matrix so that all the columns left are quantile and network variables. Normalization and centering are applied to the centralities matrix so that they have a similar scale to the quantile matrix. We then combine the quantile matrix and the rescaled network matrix to construct an extended featurization matrix, which we called the neighborhood matrix earlier.

\begin{verbatim}
    neighborhood_info_patient4 <- cbind(
        Quantiles_patient4 %>% select(-index, -n_neighbor),
        scale(centrality_patient4 %>% select(-index)))
\end{verbatim}

The final step is to input the neighborhood matrix, the cell dataset \texttt{patient4}, and the names of the variable of interest in the function \texttt{NBFvis}. This returns an interactive Shiny app that was the source of figures in Section \ref{sec:data},

\begin{verbatim}
    NBF_vis(
        matrix = neighborhood_info_patient4,
        origin_data = patient4,
        var_names = colnames(patient4)[17:57])
\end{verbatim}

\section{Discussion}
\label{sec:discussion}

We have presented a method for visualizing spatial omics datasets that integrates dimensionality reduction methods like UMAP with neighborhood-based featurization based on quantiles and network properties. 
According to the results of our simulation, dimensionality reduction based on genomic features alone has difficulty identifying microenvironments because the associated embeddings are dominated by differences in expression patterns across cell types. Also, $K$-means clustering on the UMAP embeddings from this approach results in spatially inconsistent clusters, making it difficult to identify potential microenvironments. In contrast, our approach, though simple to implement, is able to avoid these problems by leveraging neighborhood information of cells. After combining neighborhood-based statistics like quantiles and centralities, we can detect microenvironments with mixed cell types, paralleling our simulation results. Furthermore, spatially consistent $K$-means clusters can be derived, supporting discovery of microenvironments. 

We apply our methodology to the spatial omics dataset of \citep{keren2019mibi} and find five spatially continuous microenvironments in the cells' spatial plot. We compared this result with the analogous approach based on cell-level data and found that it is more difficult to identify meaningful microenviornments without an initial featurization step.

One advantage of our methodology is that the choice of neighborhood-based featurization is flexible. In our example, we use neighborhood quantiles of principal components and network statistics to build the neighborhood matrix for UMAP. These statistics could be replaced by other neighborhood-based statistics like cell-type diversity or local modularity. Also, the embedding and clustering methods are not fixed. We could use alternative dimensionality reduction methods like $t$-distributed stochastic neighbor embedding ($t$-SNE) and PCA or clustering methods like spectral clustering depending on the problem structure.

There are several avenues to develop this work. First, we treat the vertices in the neighborhood networks identically, ignoring their cell types. This is convenient for the computation of network statistics, but information is nonetheless lost. To address this, it may be possible to build neighborhood networks with different vertex types and compute corresponding network statistics. A second question is how to combine matrices. Our featurization is based on matrices from two groups of statistics (quantiles and network statistics), and their variances and interpretation could be quite different according to their groups. Is there a more principled approach to scaling and combining these measures into a single featurization? One possible solution could be Multiple Factor Analysis, which distinguishes between groups of statistics \citep{pages2014multiple}. Thirdly, we use $K$-means clustering in our methodology, which is a common choice but far from the best clustering algorithm for low-dimensional embeddings. $K$-means clustering is sensitive to outliers in the embedding plot and assumes spherical clusters, making it potentially unreliable. Spectral clustering could be a potential improvement, because it is more sensitive to the gradient structures in the UMAP embeddings.

\section*{Acknowledgements}

We would like to express our gratitude to all the members in our group, who provided us with precious suggestions for the modification of our methodology and Shiny app. In particular, we thank MinXing Zheng for suggestions on the choices of network statistics and Xinran Miao and Hanying Jiang for help improving the user interface to our Shiny app.

\bibliographystyle{unsrtnat}
\bibliography{refs}

\section*{Appendix}

Network statistics implemented in NBFvis. 

\begin{enumerate}
    \item Degree 
    \item Number of Nodes
    \item Betweenness Centrality
    \item Closeness Centrality
    \item Eigenvector Centrality Scores
    \item Eccentricity Scores
    \item Subgraph Centrality Scores
    \item Load Centrality Scores 
    \item Gil-Schmidt Power Index
    \item Information Centrality Scores
    \item Stress Centrality Scores 
    \item Average Distance
    \item Barycenter Centrality Score
    \item Latora Closeness Centrality
    \item Residual Closeness Centrality
    \item Communicability Betweenness Centrality
    \item Cross-clique Connectivity Centrality
    \item Decay Centrality
    \item Diffusion Degree
    \item Radiality Centrality
    \item Geodesic k-path Centrality
    \item Laplacian Centrality
    \item Leverage Centrality
    \item Lin Centrality
    \item Lobby Centrality
    \item Markov Centrality Score
    \item Maximum Neighborhood Component 
    \item Semi Local Centrality
    \item Topological Coefficient 
\end{enumerate}

\end{document}